\documentclass{article}

\usepackage{times}
\usepackage{graphicx} 
\usepackage{subfigure} 
\usepackage[percent]{overpic}
\usepackage{natbib}

\usepackage[draft]{hyperref}

\usepackage[accepted]{icml2017}

\icmltitlerunning{Stochastic  Bouncy  Particle Sampler}

\usepackage{mathtools}
\usepackage{amsmath}
\usepackage{amsthm}
\usepackage{amssymb}
\usepackage{graphicx}
\usepackage[small,bf]{caption}
 \usepackage{nicefrac}       
\usepackage{bm}

\usepackage{algorithm}
\usepackage{wrapfig}

\usepackage{tcolorbox}

\graphicspath{{./figures_camera/}}

\newcommand{\eq}{\begin{equation*}}
\newcommand{\en}{\end{equation*}}
\newcommand{\eqa}{\begin{eqnarray*}}
\newcommand{\ena}{\end{eqnarray*}}
\newcommand{\eqn}{\begin{equation}}
\newcommand{\enn}{\end{equation}}
\newcommand{\be}{\begin{equation}}
\newcommand{\ee}{\end{equation}}
\newcommand{\eqan}{\begin{eqnarray}}
\newcommand{\enan}{\end{eqnarray}}

\newcommand{\nn}{\nonumber}

\newcommand{\A}{ {\bf A} }

\newcommand{\w}{ {\bf w} }
\newcommand{\Dt}{ \Delta t }
\newcommand{\x}{ {\bf x} }
\newcommand{\z}{ {\bf z} }

\newcommand{\vv}{ {\bf v} }

\newcommand{\pmat}{\begin{pmatrix}}
\newcommand{\pman}{\end{pmatrix}}

\begin{document}



\twocolumn[
\icmltitle{Stochastic   Bouncy  Particle Sampler}
           
\icmlsetsymbol{equal}{*}
\begin{icmlauthorlist}
\icmlauthor{Ari Pakman}{equal,col}
\icmlauthor{Dar Gilboa}{equal,col}
\icmlauthor{David Carlson}{duke}
\icmlauthor{Liam Paninski}{col}
\end{icmlauthorlist}

\icmlaffiliation{col}{Statistics Department and Grossman Center for the Statistics of Mind, Columbia University, New York, NY 10027, USA}
\icmlaffiliation{duke}{Duke University, Durham, NC 27708, USA}

\icmlcorrespondingauthor{Ari Pakman}{aripakman@gmail.com}

\icmlkeywords{MCMC, BPS, machine learning, ICML}

\vskip 0.3in
]

\printAffiliationsAndNotice{\icmlEqualContribution} 

\begin{abstract}
We introduce a stochastic version 
of the non-reversible, rejection-free Bouncy Particle Sampler (BPS), a Markov process whose sample trajectories are piecewise linear,
to efficiently sample  Bayesian posteriors in big datasets.
We prove that in the BPS no bias is introduced by noisy evaluations of the log-likelihood gradient. 
On the other hand, we argue that efficiency considerations favor a small, controllable bias, in exchange for faster mixing. 
We introduce a simple method that controls this trade-off. 
We illustrate these ideas in several examples  which outperform previous approaches.
\end{abstract}

\section{Introduction}
The advent of the Big Data era presents special  challenges to practitioners of Bayesian modeling  because typical sampling-based inference 
methods have a computational cost per sample linear in the size of the dataset. 
This computational burden has been addressed in recent years through two major approaches (see~\cite{bardenet2015markov}  for a recent overview):
(i) split the data into batches and combine posterior samples obtained in parallel from each batch, 
or (ii) use  variants of the Markov Chain Monte Carlo (MCMC) algorithm
that only query a subset of the data at every iteration.
Our interest in the paper is in the latter approach, where many methods are based on modifying both steps of the Metropolis-Hastings (MH) algorithm:  
in the proposal step, only a mini-batch  of the data is used, 
and the accept-reject step is either ignored or
approximated~\cite{korattikara2013austerity, bardenet2014towards}. 
This strategy has been explored using proposals from Langevin~\cite{welling2011bayesian}, Riemannian Langevin~\cite{patterson2013stochastic},  Hamiltonian~\cite{chen2014stochastic} and 
Riemannian Hamiltonian~\cite{ma2015complete} dynamics.
Other relevant works include~\cite{ahn2012bayesian,ding2014bayesian}.

Despite the success  of the above approach, 
the partial  accept-reject step is a source of  bias, the precise size of which is difficult to control, and which tends to be amplified by the noisy evaluation of the gradient. 
This has motivated the search for unbiased stochastic samplers, such as the Firefly MCMC algorithm~\cite{maclaurin2014firefly}, the debiased pseudolikelihood approach of~\cite{quiroz2016exact}, and the quasi-stationary distribution approach of~\cite{pollock2016scalable}.

The present work is motivated by the idea that the bias could  be reduced by starting from a rejection-free  MCMC algorithm, avoiding thus the Metropolis-Hastings algorithm altogether. 
Two similar algorithms of this type have been recently proposed: the Bouncy Particle Sampler (BPS)~\cite{peters2012rejection,bouchard2015bouncy},
and Zig-Zag Monte Carlo~\cite{bierkens2015piecewise,bierkens2016zig}. 
These algorithms sample from the  target distribution through  non-reversible, piecewise linear Markov processes. Non-reversibility (i.e., the failure to satisfy detailed balance)
has been shown in many cases to yield faster mixing rates~\cite{neal2004improving,vucelja2014lifting,bouchard2015bouncy}.

Our contributions in this paper are twofold. Firstly, we show that the BPS algorithm is particularly well suited to sample from posterior distributions of big datasets,
because the target distribution is invariant under zero-mean noisy perturbations of the log-likelihood gradient,
such as those introduced by using mini-batches of the full dataset in each iteration. 

Stochastic variants of BPS or Zig-Zag that preserve exactly the target distribution have been proposed,
such as Local BPS~\cite{bouchard2015bouncy} or Zig-Zag with subsampling (ZZ-SS)~\cite{bierkens2016zig},
but they lead to extremely slow mixing because are based on overly conservative bounds (which moreover must be derived on a case-by-case basis, and in many cases may not hold at all).
This leads us to our second contribution, the Stochastic Bouncy Particle Sampler (SBPS), 
a stochastic version of the BPS algorithm which trades a small amount of bias for  significantly reduced variance, yielding superior performance (and requiring no parameter tuning or derivation of problem-specific bounds) compared to existing  subsampling-based Monte Carlo methods.  SBPS inherits the piecewise linear sample paths of BPS, and therefore enjoys faster convergence of empirical means, particularly of rapidly varying test functions, compared to more standard approaches.

We organize this paper as follows.
In Section \ref{BPS} we review the  Bouncy Particle Sampler, in Section \ref{resilience} we study the invariance of  the target distribution under noise perturbations to the BPS updates, in Section \ref{SBPS} we introduce SBPS, 
and in Section \ref{prec_BPS} a preconditioned variant.
In Section~\ref{related_works} we discuss related works and in Section \ref{examples} we illustrate the advantages of SBPS in several examples.

\section{The Bouncy  Particle Sampler}
\label{BPS}
 \begin{figure*}[t!]
  \centering
\begin{tabular}{lcc}
  \includegraphics[width=.30\textwidth]{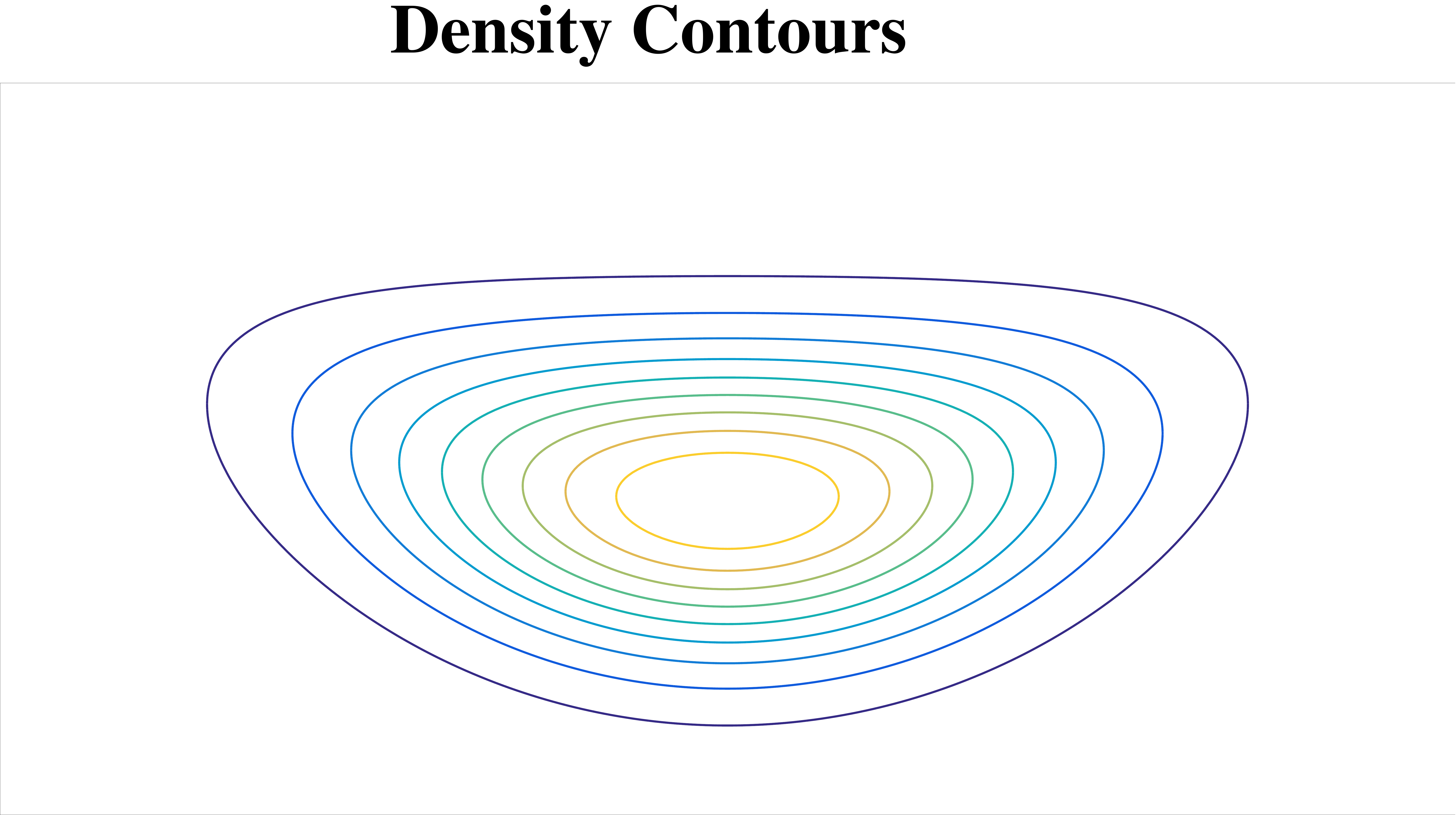} &   \includegraphics[width=.30\textwidth]{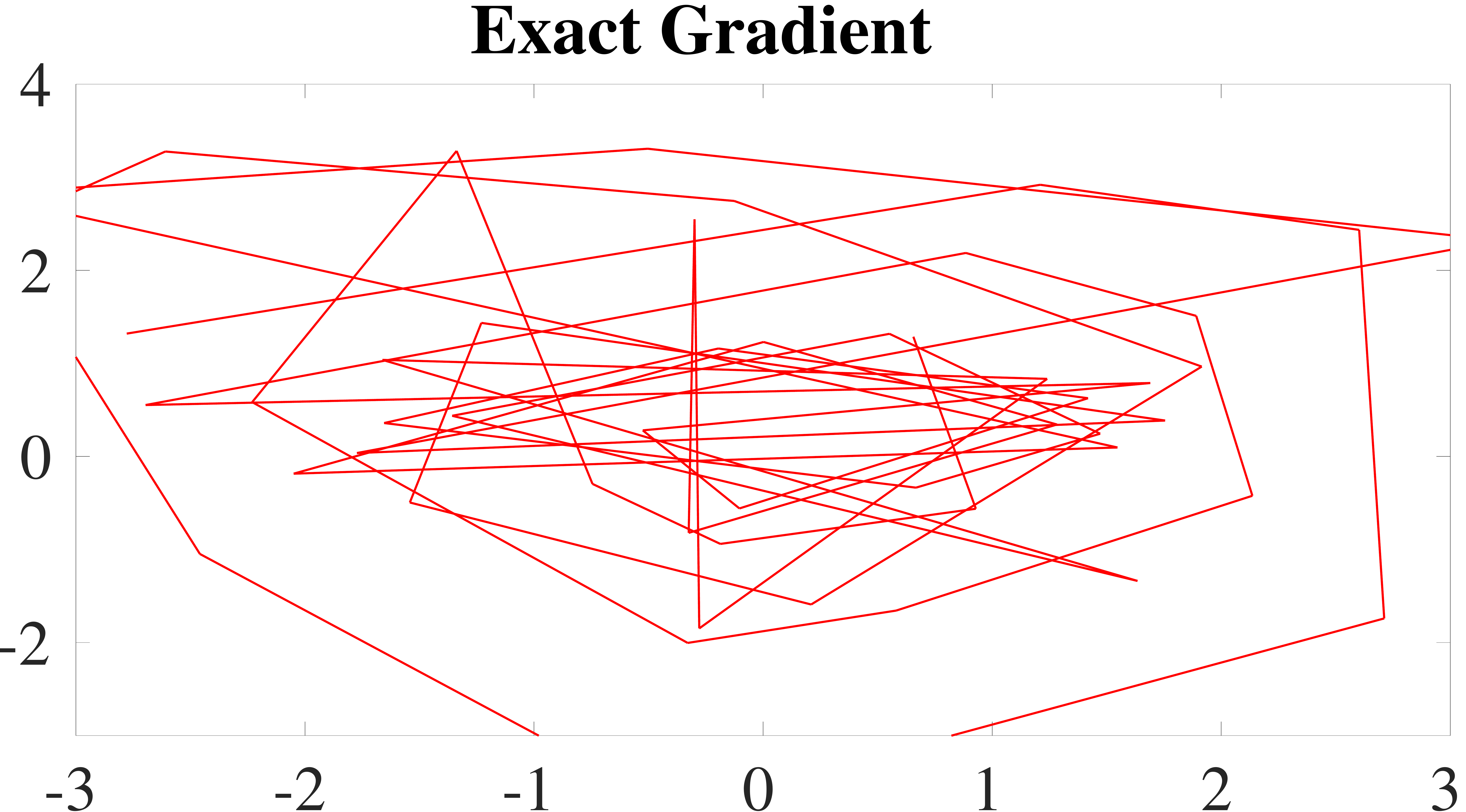} &    \includegraphics[width=.30\textwidth]{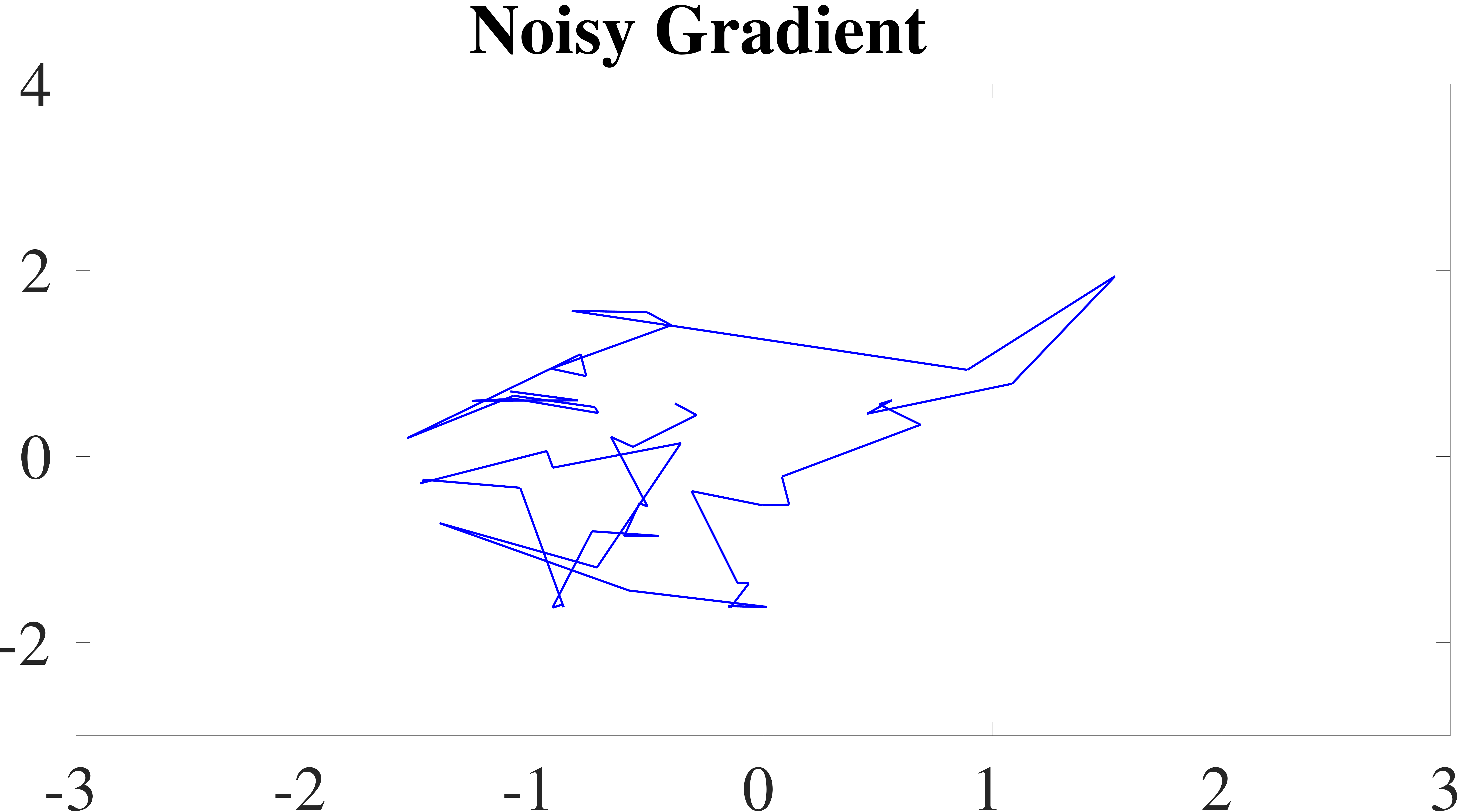} 
  \\
  \includegraphics[width=.30\textwidth]{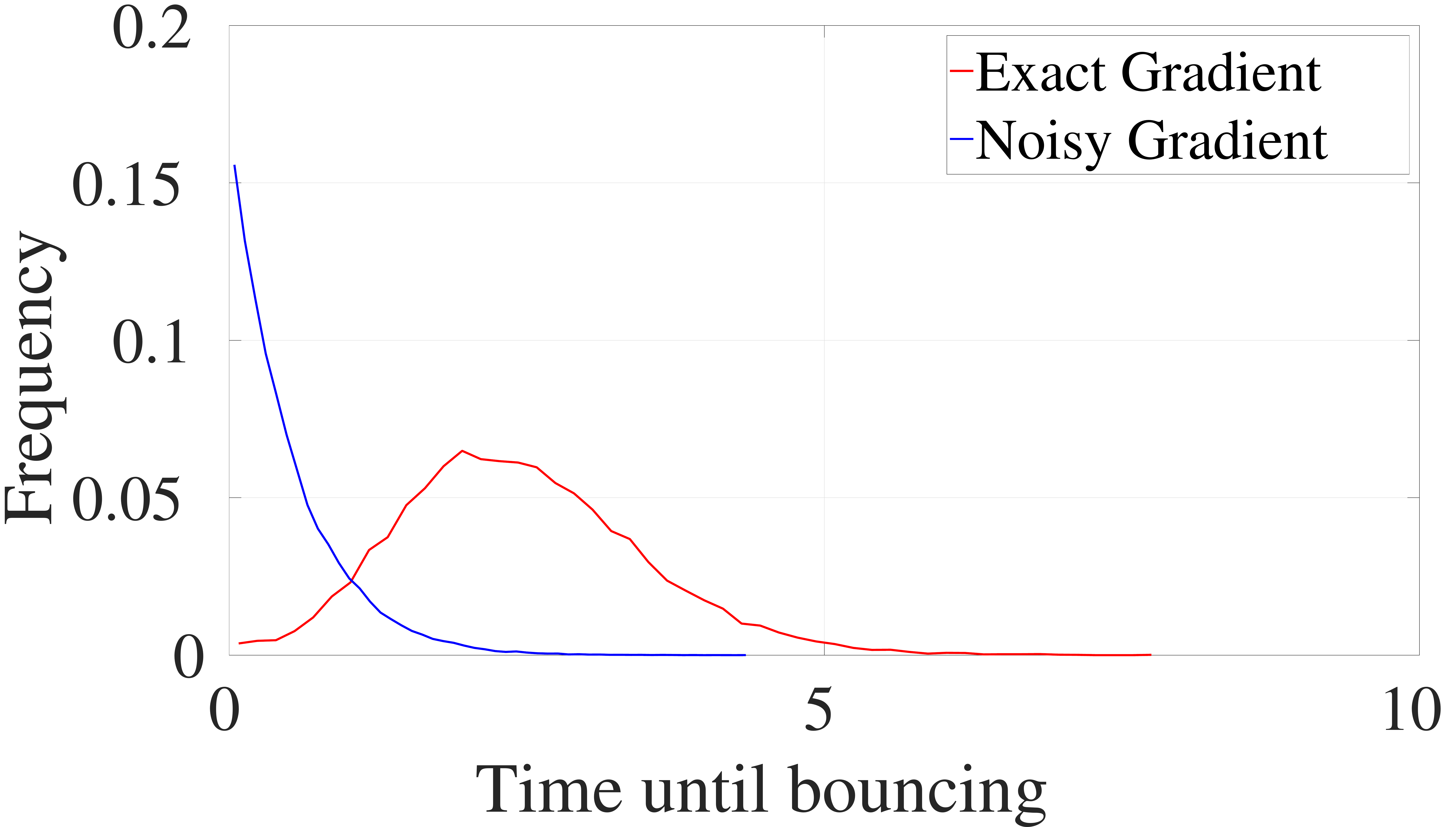}       &   \includegraphics[width=.30\textwidth]{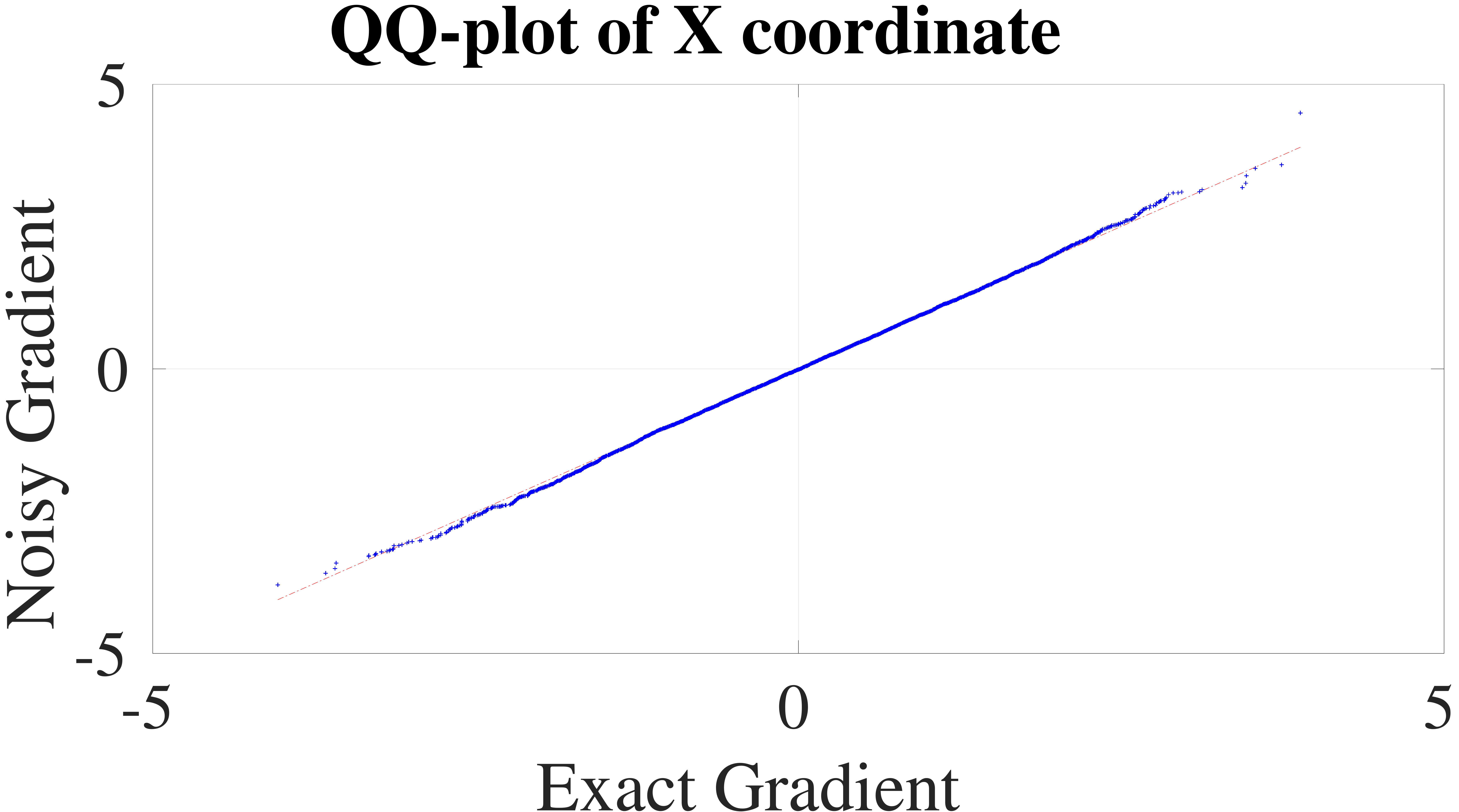} &   \includegraphics[width=.30\textwidth]{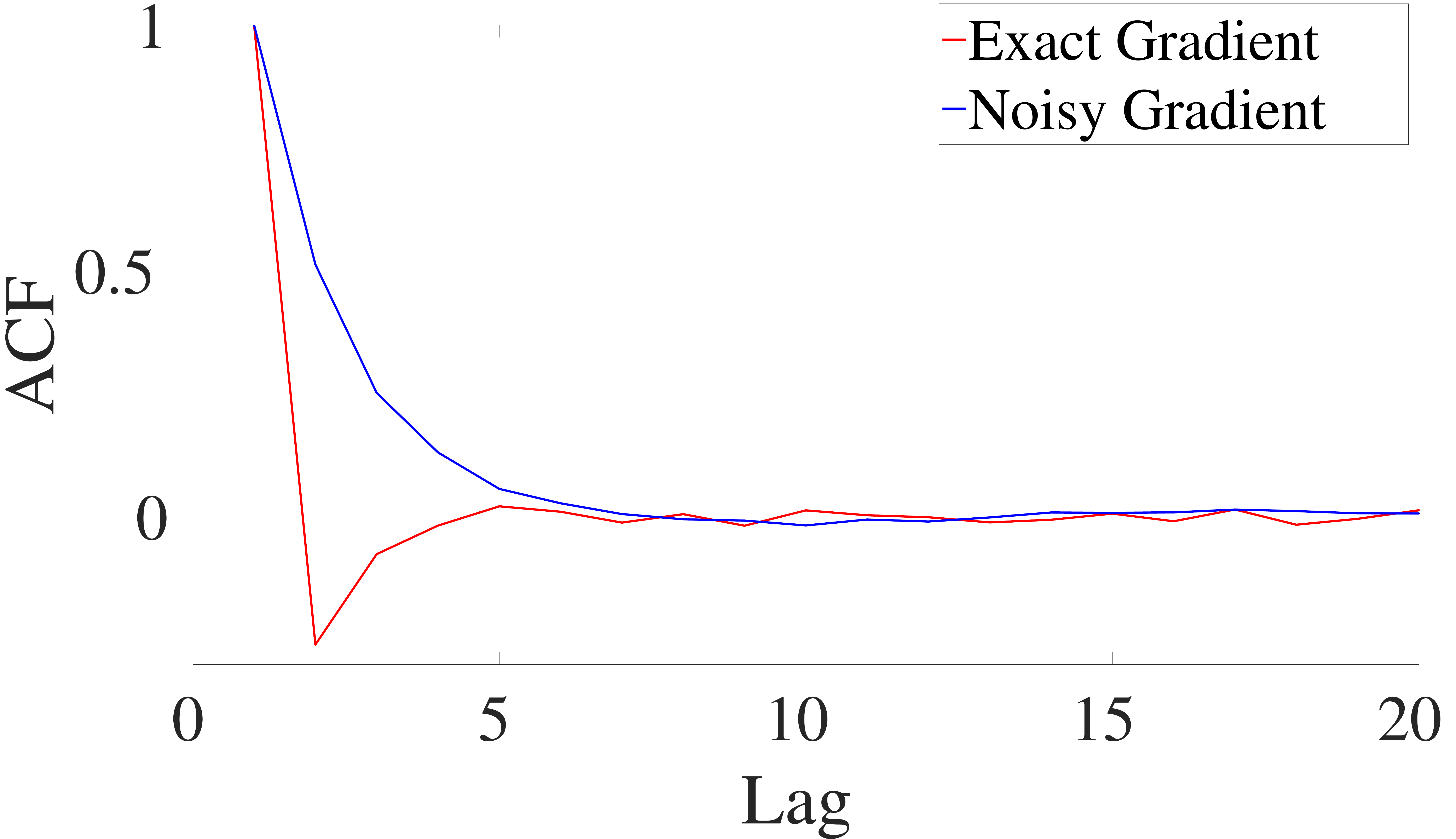}
\end{tabular}
     \caption{Noisy vs. noiseless gradients in BPS. \textit{Above}: Contour plot of the 2D density considered and sample BPS trajectories during 50~bounces, with exact  and noisy  gradients.
     The noise was sampled in each component from a ${\cal N}(0, 5^2)$ distribution. 
     \textit{Below, Left}: smoothed histogram of travel times until bouncing, with shorter times for noisy gradients.
     \textit{Middle}: QQ-plot of one of the coordinates, showing that the invariant distribution is not changed by the noise. 
     \textit{Right}: ACFs of one of the  coordinates, with slower mixing per iteration in the noisy case. 
     However, note that these ACF plots do not account for computational cost per iteration.
     }
\label{noise_bps}
\end{figure*}

\begin{algorithm}[t!]
  \begin{tcolorbox} 
  \caption{Bouncy Particle Sampler}
  \label{bpsalgo}  
      \begin{algorithmic}
	  \STATE Initialize particle position $ \w_0 \in \mathbb{R}^D$ and  velocity $\vv \in S^{D-1}$ 

	  \WHILE{$\text{desired}$}
		  \STATE Sample Poisson process first arrivals $t_{r}, t_{b}$ with rates $\lambda_r$,$\lambda(t) = [\vv \cdot \nabla U(\w_0 + \vv t)]_+$ 
		  \STATE Let $t = \min(t_{b},t_{r})$ 	
		  \STATE 	Move $\w_{t} = \w_0 + \vv t \,,$
		  \IF {$t_{b}<t_{r}$}	 
			  \STATE Reflect 
			  $\vv \leftarrow  \vv -2\frac{(\vv \cdot \nabla U(\w_{t})) \nabla U(\w_{t}) }{||\nabla U(\w_{t})||^2}$

		  \ELSE 
			  \STATE Refresh: sample  $\vv \sim \textrm{Unif}[S^{D-1}]$ 
		  \ENDIF
		  \STATE Let $\w_0 \leftarrow \w_{t}$ 
	  \ENDWHILE
	  \STATE RETURN  piecewise linear trajectory of $\w$
      \end{algorithmic}
  \end{tcolorbox}
\end{algorithm}

Consider a  distribution $p(\w) \propto e^{-U(\w)} \,,  \w \in \mathbb{R}^D$,
where the  normalization factor  may be intractable. 
The Bouncy Particle Sampler (BPS), proposed in~\cite{peters2012rejection,monmarche2014piecewise} 
and formalized and developed in~\cite{bouchard2015bouncy}, 
introduces a random velocity vector  $\vv$ distributed uniformly in the unit sphere $S^{D-1}$, 
and defines a continuous Markov process in $(\w, \vv)$.  To describe this process we begin in discrete time and then take  the continuous-time limit. 
Denoting time by $t$, consider a discrete-time Markov process that acts on the variables $(\w, \vv)$ as
\eqan 
(\w, \vv)_{t + \Delta t} = 
\left\{ \!\!\!\!\!
 \begin{array}{rl}
 (\w \! + \! \vv \Delta t  , \vv)   & \!\!\!\!\! \textrm{w/prob. }   1-\Dt [ G]_+ 
\\
(\w \! + \! \vv \Delta t , \vv_r) &  \!\!\!\! \textrm{w/prob. }   \Dt [G]_+ 
\end{array}
\right.  
\label{pyse}
\enan 
where 
\eqan 
[x]_{+} = \max(x,0) \,,
\\
G =  \vv \cdot \nabla U(\w) \,,
\label{dG}
\\
\textstyle \vv_r =  \vv -2\frac{(\vv \cdot \nabla U(\w)) \nabla U(\w) }{||\nabla U(\w)||^2} \,.
\label{vvr}
\enan 
Note that $G$ in (\ref{dG}) is the directional derivative of $U(\w)$ in the direction $\vv$,
and $\vv_r$ is  a reflection of $\vv$ with respect to  the plane perpendicular to the gradient $\nabla U$,
satisfying  $\vv_r \cdot \nabla U =-\vv \cdot \nabla U$ and $(\vv_r)_r = \vv$. 
In other words, the particle $\w$ moves along a straight line in the direction of $\vv$ and this direction is reflected as (\ref{vvr}) with probability $\Dt [G]_+$.
This probability is non-zero only if the particle is moving in a direction of lower target probability $p(\w)$, or equivalently higher potential~$U(\w)$.

Applying the transition (\ref{pyse}) repeatedly and taking $\Delta t \rightarrow 0$, the random reflection point becomes 
an event in an inhomogeneous Poisson process with intensity  $[G]_+$.
The resulting sampling procedure generates a piecewise linear Markov process~\cite{davis1984piecewise,dufour2015numerical}, and
is summarized in Algorithm~\ref{bpsalgo}.
Note that the algorithm also includes occasional resamplings of $\vv$,  to ensure ergodicity~\cite{bouchard2015bouncy}.
Remarkably, in the limit $\Delta t \rightarrow 0$,  the algorithm leaves the joint factorized distribution $p(\w)p(\vv)$ invariant, 
as we review in Supp.\ Material~A.1.

The Zig-Zag process \cite{bierkens2015piecewise,bierkens2016zig} is similar to BPS, 
but velocity components can take only $\pm 1$  values, and
the piecewise linear trajectories change direction only in a {\it single} coordinate at each random breakpoint. 
For a review of these methods, see~\cite{fearnhead2016piecewise,bierkens2017piecewise}.

\section{Noise Resilience and Big Data}
\label{resilience}

\subsection{Noise Resilience}

Let us assume that only  a noisy version of the gradient  
is available to compute the probability of bouncing  and the reflected velocity in (\ref{vvr}).
In the Big Data scenario described below, this  is the result of using a random subset of the data at each gradient evaluation,  
and can be represented as
\eqan 
\nabla \tilde{U}(\w) = \nabla U(\w) + {\bf n}_{\w} \,, \qquad  {\bf n}_{\w} \sim  p(  {\bf n}_{\w} |\w)\,,
\label{unoise}
\enan 
where ${\bf n}_{\w} \in {\mathbb R}^D$ and  $p(  {\bf n}_{\w} |\w)$ has zero mean.
\vskip .4cm
\noindent
{\bf Theorem 1:} {\it The invariance of $p(\w,\vv)$ under the BPS algorithm is unaffected by the zero-mean noise~(\ref{unoise}) if ${\bf n}_{\w_1}$ and ${\bf n}_{\w_2}$ 
are independent for $\w_1 \neq \w_2$. } 
\vskip .4cm
\noindent
See Supp.\ Material A.2 for a proof sketch. 
Defining $\tilde{G} =  \vv \cdot \nabla \tilde{U}(\w)$,   the intensity of the inhomogeneous Poisson process~$[\tilde{G}]_+$, which determines the time of the velocity bounce,
now becomes stochastic, and the resulting point process is called a doubly stochastic, or Cox, process~\cite{cox1955some,grandell1976doubly}.
The effect of the gradient noise is to increase the average point process intensity, since $E\left[ [\tilde{G}]_+ \right] \geq \left[ E[\tilde{G}] \right]_+$, from Jensen's inequality.
This leads to more frequent bounces and typically a slower mixing of the Markov process, as
illustrated in Figure~\ref{noise_bps}. 

Many Cox processes  are based on Poisson intensities obeying  stochastic differential equations, 
or assume that the joint distribution at several $\w$'s has a non-trivial $\w$-dependent structure.
Our case is different because we assume that ${\bf n}_{\w_1}$ and ${\bf n}_{\w_2}$ are independent even when $\w_1$ and $\w_2$ are infinitesimally close.


\subsection{Sampling from Big Data posteriors}
In a prototypical Bayesian  setting,  we have a prior $f(\w)$, i.i.d.~data points $x_i$, and the negative log-posterior gradient is  
\begin{equation}
\textstyle
\nabla U(\w)= -\nabla\left[\log f(\w)+\underset{i=1}{\overset{N}{\sum}}\log p(x_{i}|\w) \right] \,.
\end{equation}
When $N$ is big we consider replacing the above gradient by the noisy approximation
\begin{equation}
\textstyle
\nabla \tilde{U} (\w)= -\nabla\left[\log f(\w) +\frac{N}{n}\underset{i=1}{\overset{n}{\sum}}\log p(x_{r_i}|\w)\right] \,,
\label{tU}
\end{equation}
where $n  \ll N$ and the $n$ indices $\{ r_i \}$ are sampled randomly without replacement.
To sample from the posterior using the noisy gradient (\ref{tU}),
we want to simulate the first arrival time in a doubly stochastic Poisson process with random intensity $[\tilde{G}(t)]_+$, where
\eqan 
\tilde{G}(t) = \vv \cdot \nabla \tilde{U}(\w+ \vv t) \,.
\label{Gt}
\enan 
Note that  $\tilde{U}$ is a stochastic process, 
and noise independence for different  $\w$'s implies that different  $t$'s require independent mini-batches.
Out of several methods to sample from (noisy) Poisson processes, 
the thinning method~\cite{lewis1979simulation} is compatible with the noise independence assumption.
This is a  form of rejection sampling which proposes a first arrival  time $t$,  sampled from an 
inhomogeneous Poisson process with intensity $\lambda(t)$ such that 
$\lambda(t) \geq  [\tilde{G}(t)]_+$
The particle moves a distance $t \vv$, and accepts the proposal to bounce the velocity with probability $[\tilde{G}(t)]_+/\lambda(t)$.
Note that this accept-reject step is different from the MH algorithm~\cite{robert2013monte}, since the particle always moves the distance $t \vv$, 
and a rejection only affects the velocity bouncing.  This can greatly improve the efficiency of the sampler. 
As in the noiseless case, one should in general also resample  $\vv$ occasionally, to ensure ergodicity~\cite{bouchard2015bouncy},
although in the examples we considered this was not empirically necessary, since the  mini-batch noise
serves to randomize the velocity sufficiently, preventing ``non-ergodic'' trajectories that do not explore the full space.

In some special cases one can derive a bound $\lambda(t)$ that always holds~\cite{bouchard2015bouncy,bierkens2017piecewise}.
But this is atypical, due to the dependence of $\tilde{G}(t)$ in (\ref{Gt}) on the changing velocity $\vv$ and the mini-batch noise. 
Even when such bounds do exist, they tend to be conservatively high, 
leading to an inefficient sampler with many rejected proposals (wasting many mini-batches of data) before accepting.

Instead, we propose below an adaptive approximate bound 
which achieves a bias-variance trade-off between 
the frequency of the bounce proposals
and a controllable probability of bound violation. 

\section{Proposal  from Local Regression } 
\label{SBPS}

Our approach to an adaptive and tractable  proposal intensity $\lambda(t)$ 
relies  on a predictive model of~$\tilde{G}$ based on previous observations; 
the key idea is to exploit the correlations between nearby $\tilde{G}$ values.
The upper value of the resulting predictive confidence band can then be used as $\lambda(t)$, and this band is adaptively updated as more proposals are generated.

While there are many possibilities for such a predictive model, we found that a simple local linear model was very effective and computationally trivial.
Consider then the linear regression  of $m$ observed values  $\tilde{G}_i \equiv \tilde{G}(t_i)$ since the previous bounce,
\eqan 
\tilde{G}_i  = \beta_1 t_i + \beta_0 + \varepsilon_{t_i}  \qquad \varepsilon_{t_i} \sim N(0, c^2_{t_i})    \,,
\label{noisyG}
\enan 
where $i=1, \ldots, m$ and the noise variance can be estimated from the mini-batch in~(\ref{tU}) as
\eqan 
\textstyle
c_t^2 = \frac{N^2}{n} (1-\frac{n}{N})\textrm{Var}_i  \left[ \vv \cdot  \nabla \log p(x_{r_i}|\w)  \right ]  \,.
\label{mb_var}
\enan 
Here  $\textrm{Var}_i$ denotes the sample variance of the mini-batch, and we included the {\it finite population correction factor} $(1-\frac{n}{N})$
because the indices $\{ r_i \}$ are sampled without replacement. 
The Gaussian noise assumption in $\tilde{G}(t)$ in (\ref{noisyG}) is valid 
when the mini-batch is sufficiently large that we can appeal to a central limit theorem.  (For heavy-tailed noise we could consider more robust estimators, but we do not pursue this direction here.)

Adding a Gaussian prior ${N}(\mu,\sigma^2)$ to $\beta_1$, and defining  $\mathbf{x}_{i} \equiv (1,t_{i})$, the log posterior of  $\bm{\beta} = (\beta_0, \beta_1)^T$ is  
\eqan 
\textstyle
2 \log p(\bm{\beta}|\{t_{i},\tilde{G}_{i},c^2_{t_i}\}) &=&\textstyle -\underset{i=1}{\overset{m}{\sum}} \frac{ (\tilde{G}_{i}-\mathbf{x}_{i}\cdot \bm{\beta})^{2}}{c_{t_i}^{2}} 
\nn
\\
&-&\textstyle \frac{(\beta_{1}-\mu)^{2}}{\sigma^{2}} + const.
\nn
\enan
Let $\hat{\bm{\beta} }$ and $\bm{\Sigma}$ be the mean and covariance of this distribution. 
Using these estimates, we obtain the predictive distribution $\hat{G}(t)$ for $\tilde{G}(t)$ for $t> t_m$, 
\eqan
\hat{G}(t) = \hat{\beta}_1 t + \hat{\beta}_0  + \eta_t   \qquad  \eta_t \sim N(0, \rho^2(t) )
\label{Gpred}
\enan
\eqan 
\text{where}\quad \rho^2(t) = \mathbf{x} \bm{\Sigma} \mathbf{x}^T  + c^2_{t_m}
\enan 
with $\x = (1,t)$. Note that as usual the noise variance is different in (\ref{noisyG}) and (\ref{Gpred}), since in (\ref{noisyG}) we are fitting 
observed pairs $\tilde{G}_i, t_i$, while in (\ref{Gpred}) we are predicting the value of $\tilde{G}(t)$ and we include 
 the uncertainty from the $\hat{\bm{\beta} }$ estimates.  Also, for simplicity we extrapolate the observation noise to be the same as  in the last mini-batch,~$c^2_{t_m}$. 

 We can now construct a tractable approximate thinning proposal intensity by choosing a  confidence band multiple $k$, 
and defining $\gamma(t)$ as a linear interpolation between selected points along the non-linear curve 
\eqan
\hat{\beta}_{1}t+\hat{\beta}_{0}+k \rho(t)  \,.
\label{upper_bound}
\enan
The proposal intensity is now $\lambda(t) = [\gamma(t)]_+ $, 
and sampling from an inhomogeneous Poisson process with piecewise linear rate $\lambda(t)$ can be done analytically
using the inverse CDF method. 
When a bounce time is proposed at time $t$, the particle moves a distance $t \vv$,
a noisy observation $\tilde{G}(t)$ is made as in (\ref{Gt}) and the bounce time is 
accepted with probability $\min(1,[\tilde{G}(t)]_+/\lambda(t))$.
If the bounce  is accepted, the velocity is reflected as in (\ref{vvr}) (using $\tilde U$ instead of $U$),
and the set of observed values is reinitialized with $(-\tilde{G}(t), c_{t})$,
which are the values one would obtain from sampling the same mini-batch {\it after} the bounce, since
$\vv_r \cdot \tilde U = - \vv \cdot \tilde U = -\tilde{G}(t)$.
On the other hand, if the proposal is rejected, the observed $(\tilde{G}(t), c_{t})$ are 
added to the set of observed values.
The hyperparameters $\mu, \sigma^2$ of the regression model can be learned by performing, after each bounce, 
a  gradient ascent step on the marginal likelihood, $p(\{ \tilde{G}_i \}|\mu,\sigma^2)$; this gradient can be computed analytically and does not significantly impact the computational cost. 

\begin{figure}[t!] 
  \centering
   \includegraphics[width=1\columnwidth]{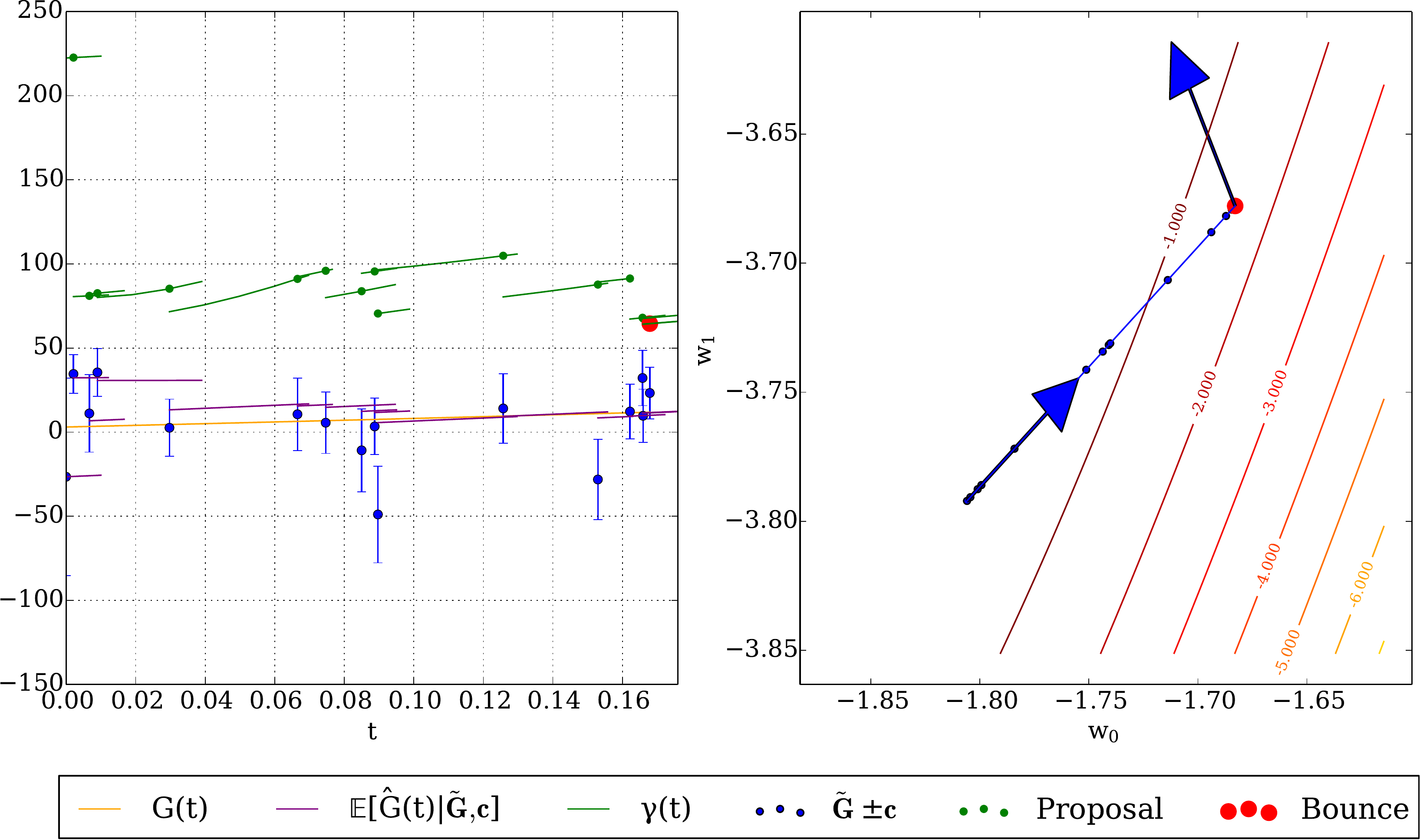}
     \caption{Thinning proposal intensity for bounce times from a linear regression predictive confidence interval applied to a two-dimensional logistic regression posterior with $N=1000,n=100$. \textit{Left:} Starting from $t=0$, the piecewise linear intensity $\gamma(t)$ is used to propose bounce times (green points). As these proposals are rejected additional observations $\tilde{G}_i$ are made until a proposal is accepted (red point). The decrease in the slope of $\gamma(t)$  indicates the decreasing 
     uncertainty in the estimated regression parameters as observations increase; note that the linear approximation for the true $G(t)$ is quite accurate here. Note also the reduced observation frequency at lower values of $G(t)$ indicating more efficient sampling than is achievable with the constant and much higher bounds used in \cite{bouchard2015bouncy,bierkens2016zig},  which were  in the range $[10^4,2*10^4]$ for this data. \textit{Right}: The corresponding SBPS particle trajectory, with arrows indicating the initial velocity and the velocity after the bounce. The contours show the Laplace approximation of the log posterior.} 
\label{SBPSfig}
\end{figure}

The linear model for $\tilde{G}$ is good when the target distribution can be locally approximated by a Gaussian, since $\tilde{G}(t)$ 
in (\ref{Gt}) is a projection of the derivative of the negative log posterior.
When the posterior is highly non-Gaussian, a decaying weight can be used for more-distant observations, leading to a local regression; 
the scale of this decay can be fit again via stochastic gradient ascent on the predictive likelihood. 
We have also explored a Gaussian Process regression model, but it did not improve over the linear model in the cases we considered.  
In Supp.\ Material~E we discuss a potential problem with our approach in the case of multimodal distributions, and propose a solution for such cases.

Finally, note that the directional derivative of $\tilde{U}(\w)$ needed in (\ref{Gt}) can in many cases be  computed at a cheaper cost 
(by a factor of $d=\dim(\w)$) than the full gradient. The latter is only needed when a bounce is accepted.  
This is in contrast to other gradient based samplers which require the full gradient at every step.

We dub this  approach to BPS with noisy gradients  Stochastic BPS (SBPS). See Supp.\ Material C  for pseudocode.
Figure \ref{SBPSfig}  illustrates the evolution of these dynamic proposal intensities in a simple example.
In Section 5, we consider a variant to SBPS, called pSBPS, 
that learns a diagonal preconditioning factor for the gradient, and leads to a more efficient
exploration of the space when the posterior is highly anisotropic and roughly axis-aligned.

\subsection{Bias in the Samples  }
The constant $k$ in (\ref{upper_bound}) controls the tradeoff between bias from 
possible $[\tilde{G}(t)]_+/\lambda(t)>1$ cases and lower computational cost: higher $k$ leads to a more conservative (higher) proposal intensity and therefore a less-biased but more data-inefficient sampler.
We present a bound on the Wasserstein distance between the exact and bias distributions in Supp.\ Material~B, 
and explore this bias-variance tradeoff further in  Supp.\ Material~F.
A quick bias diagnostic is the rate at which the bound is violated, i.e., cases with 
$[\tilde{G}(t)]_+/\lambda(t) >1$; if this rate is significantly higher than expected under the local linear regression model, then a different approach should be considered.

\section{Preconditioned SBPS}
\label{prec_BPS}
Consider now the linear transformation $\w = \A\z$
with an arbitrary square matrix $\A$.  A distribution $p(\w)$ of interest can be expressed in terms of $\z$ as
\eqan 
p_z(\z)d\z &=& p(\w(\z))d\w = p(\A\z) |\A|d\z \,,
\\
&=& \exp(-U_z(\z)) d\z \,.
\enan 
The SBPS algorithm can be applied to the density $p_z(\z)$ using the gradients of $U(\w)$. For this  note that 
$\nabla_z U_z(\z) = \A \nabla_w U(\w)$.
The Poisson intensity to compute bounces is $[G]_+$, with $G =  \vv \cdot \A \nabla U(\w)$, 
and the velocity reflection is computed as 
\eqan 
\vv_r =  \vv -2\frac{(\vv \cdot \A \nabla U(\w))\A \nabla U(\w) }{||\A \nabla U(\w)||^2} \,.
\label{vvrA}
\enan 
The piecewise linear trajectory $\z_t = \z_0 + \vv t$ becomes
$\w_t = \w_0 + \A \vv t$.
The matrix $\A$ is called a preconditioner in the optimization literature, but can also be used in a sampling context to reduce anisotropy of posterior distributions; it is often the case that a good preconditioner is not known in advance but is instead learned adaptively~\cite{duchi2011adaptive}. 

\begin{figure}[t!] 
  \includegraphics[width=0.98\columnwidth]{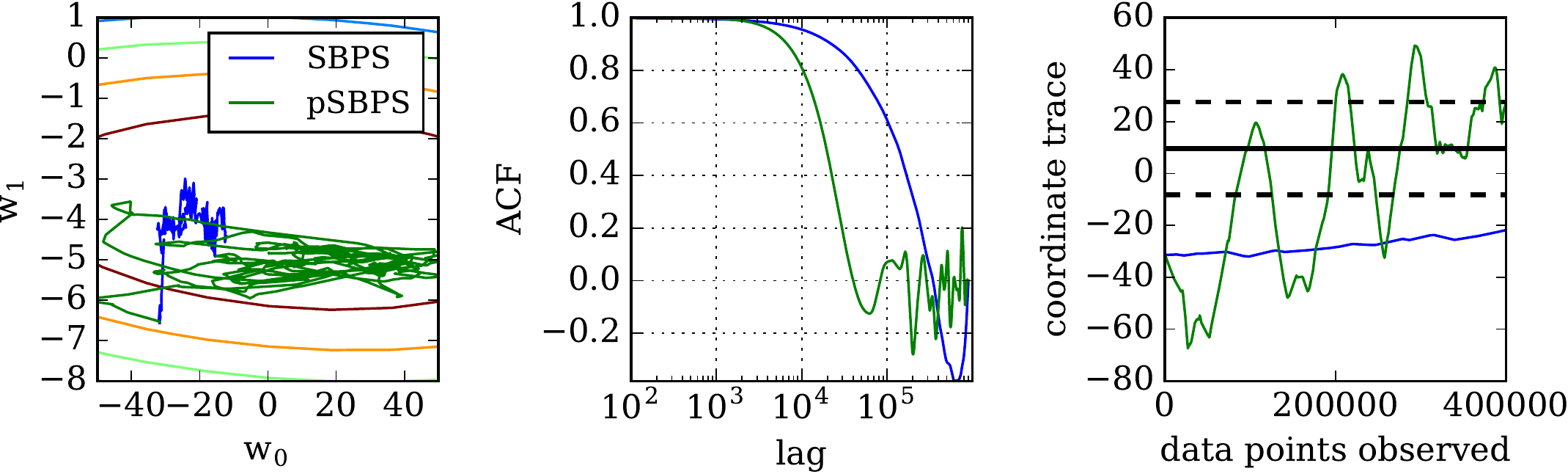}  
       \caption{Effect of diagonal preconditioning on SBPS performance. Sampling is from the logistic regression posterior as described in Section~\ref{regr} , with $d=20, N=1000, k=3, n=100$. The preconditioner parameters are $\beta=.99,\epsilon=10^{-4}$.
       Left: Contour plots of posterior log likelihood under the Laplace approximation. Center, right: ACF and trajectories in the direction of greatest covariance.}
\label{prec_fig}
\end{figure}

We use a diagonal preconditioner for simplicity. 
Denoting the $i$th component at the $j$th evaluation of the gradient by $g_i^j $, we define 
\eqan
\textstyle a_i^j&=&\textstyle\beta (g_i^j)^2 + (1-\beta)a_i^{j-1},
\\
\textstyle \tilde{a}^j&=&\textstyle\frac{1}{d}\sum_{i=1}^d\frac{1}{\sqrt{a_i^j}+\epsilon} \,,
\enan
for some $0\leq\beta\leq1,\epsilon\ll1$. 
The preconditioner at iteration $j$ is defined as 
$\A^j= \textrm{Diag} \left(\frac{\tilde{a}^j}{\sqrt{a_i^j}+\epsilon} \right)$.
This is the same preconditioner used in ~\cite{pSGLD_AAAI2016}, up to the $\tilde{a}^j$ factor; the latter is needed here in order to prevent scaling of $\tilde{G}$. 


As noted in~\cite{pSGLD_AAAI2016}, a time dependent preconditioner requires adding a term proportional to $\frac{\partial \A^j}{\partial \w}$ to the  gradient, 
yet this term is negligibly small and can be ignored when $\beta \approx 1$, since in this parameter regime the preconditioner changes slowly as a function of $j$ and thus of~$\w$.

We call this preconditioned variant pSBPS. 
It performs favorably compared to SBPS when the posterior is anisotropic and axis-aligned, since we use a diagonal approximation of the Hessian in the preconditioner. 
See~\cite{bierkens2017piecewise} for a related approach.
As Figure~\ref{prec_fig} shows, pSBPS converges to the posterior mode faster than SBPS, and mixes faster in the direction of greatest 
covariance.\footnote{pSBPS  code  at https://github.com/dargilboa/SBPS-public.}

\section{Related Works}
\label{related_works}

\begin{figure*}[t!] 
\centering
    \begin{overpic}[width=1\textwidth]{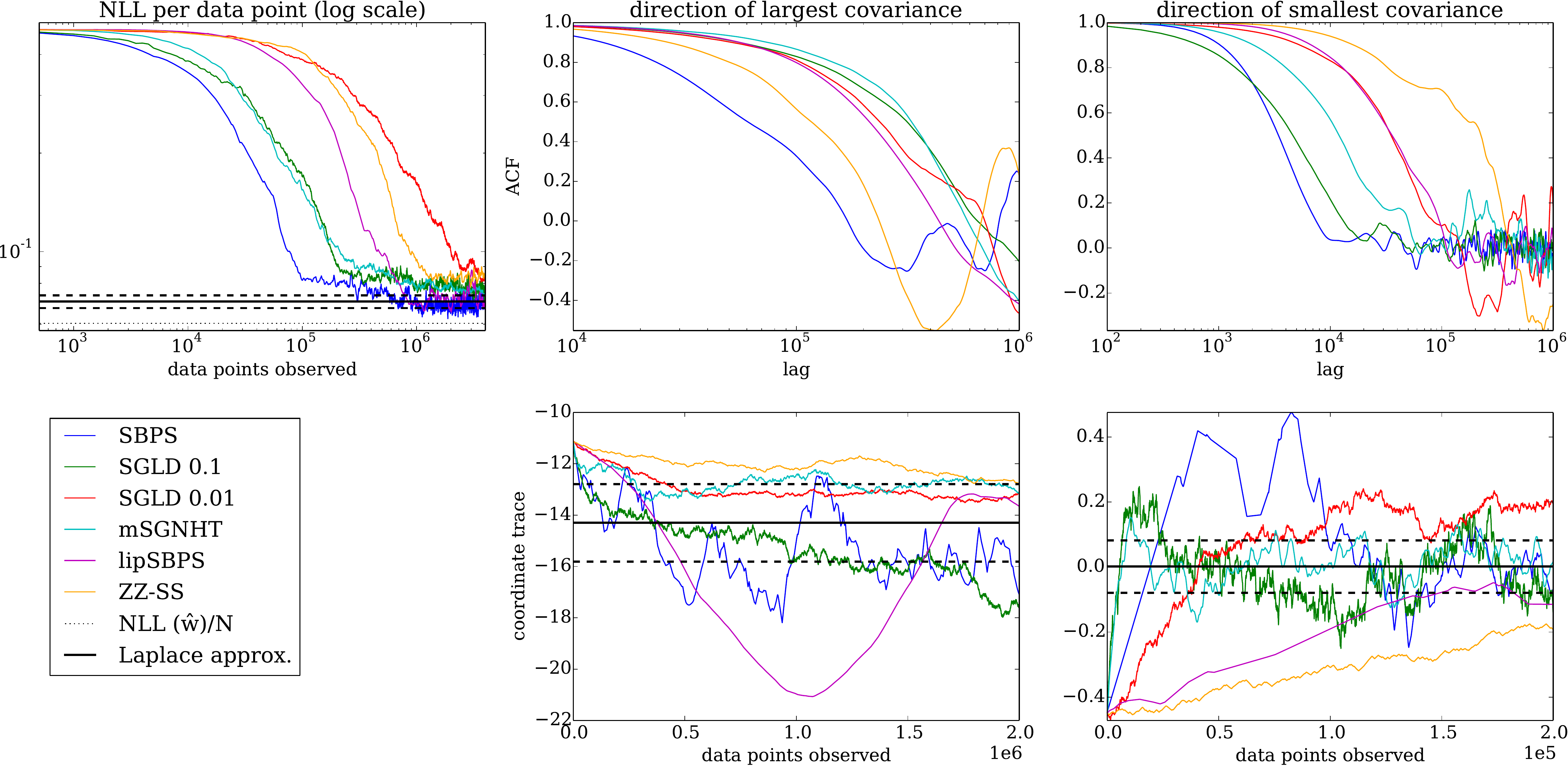}
     \put(37.5,28){\includegraphics[scale=.5]{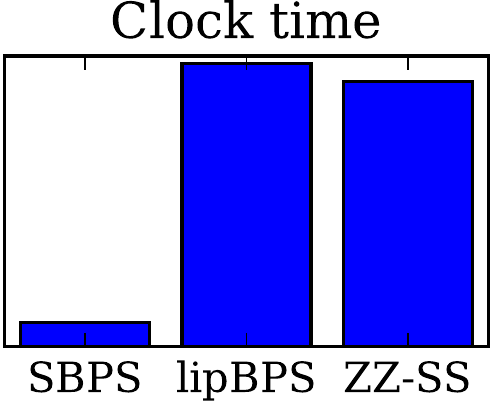}}
  \end{overpic}

\caption{Logistic regression posterior sampling, with $d=20, \, N=1000, \, k=3, \, n=100$ (best seen in color). 
\textit{Top Left:} Negative Log Likelihood (NLL) per data point of samples from SBPS compared with 
SGLD (step sizes $0.01, 0.1$), mSGNHT (step size $0.1$), lipSBPS 
and SS-ZZ (see text for definitions and references), all initialized at the same random positions. 
Also shown are the normalized NLL of the MAP estimator $NLL(\hat{\w}) / N$ and the mean $\pm$ std. dev.  of the 
Laplace approximation NLL (distributed as $\frac{1}{2N}\chi^{2}(d)+ NLL(\hat{\w}) / N$).
The continuous samplers (SBPS, SS-ZZ, lipSBPS) were run to match the data cost of the discrete (SGLD, mSGNHT),
and for their ACFs we discretized the continuous paths uniformly to obtain the same number of samples. 
Note that SBPS is the fastest to converge. 
\textit{Center/Right:} 
Trajectories and ACFs in the directions of largest  and smallest  eigenvalues of the Laplace approximation inverse Hessian. 
The ACFs were calculated after burn-in, while the trajectory plots only show initial convergence.
\textit{Inset:} CPU runtime for 100 epochs, showing a $\times 35$ advantage of $n=100$ SBPS over $n=1$ SS-ZZ and lipSBPS} 

\label{figNLL2SGLD}
\end{figure*}


{\bf Biased Samplers:}
Many stochastic gradient  samplers (e.g. \cite{welling2011bayesian}) 
can be formulated exactly using a Wiener process~\cite{ma2015complete}, 
but they are biased because (i) the Gaussian assumption in the noise may not hold for small mini-batches,  and 
(ii) the MH correction to the time discretization is avoided or approximated.
Recently, irreversible samplers have been studied in this context~\cite{ma2016unifying}.  Choosing the step size in these samplers can be quite challenging, as discussed below: too-large step sizes increase the bias, 
while too-small step sizes slow the mixing, and in generic high-dimensional examples there is no way to automatically tune the step size (though see \cite{giles2016multilevel} for recent progress).
In contrast, the bias in SBPS, controlled by the constant~$k$, does not come from time discretization, 
but from easy-to-track violations of the thinning bound when $[\tilde{G}(t)]_+/\lambda(t)>1$.

{\bf Exact non-BPS-like Samplers:}
Firefly MCMC~\cite{maclaurin2014firefly} 
augments the target distribution with one binary variable per data point, and
yields unbiased samples while only querying a subset of data points at each  iteration.
But it needs distribution-dependent  lower bounds on the likelihood and requires an initial full sweep of the data.
Also mixing can be extremely slow~\cite{quiroz2015speeding,bardenet2015markov}, and  all the dataset must be available for access 
all the time. 

Two recent novel proposals are~\cite{quiroz2016exact}, based on debiased pseudolikelihood combined with variance reduction techniques,
and~\cite{pollock2016scalable}, based on quasi-stationary distributions.
These methods are relatively more complex, and we have not yet systematically compared them against SBPS.

{\bf Exact BPS-like Samplers:} Two subsampling variants of BPS which preserve the exact distribution are 
Local BPS~\cite{bouchard2015bouncy}, that needs a pre-processing step of computational cost $O(N\log N)$,
and ZZ-SS~\cite{bierkens2016zig}. In these approaches, 
the requirement to preserve the distribution exactly  leads to extremely conservative thinning bounds, which in turn yield a very slow exploration of the 
space, as we will see below. Also, the bounds need to be rederived for each new model (if possible at all), 
unlike SBPS which can be used  for any differentiable posterior distribution.

\section{Experiments}
\label{examples}

\subsection{Logistic Regression}
\label{regr}
Although simpler MCMC methods perform well in Bayesian logistic regression (BLR) models \cite{chopin2015leave}, we begin with this well-understood
case for  comparing  SBPS against a few of the existing stochastic MCMC methods discussed in the previous section.
To generate the data, we sampled  the components of the true  $\w\in\mathbb{R}^d$ from  $\textrm{Unif}[-5,5]$ and $N$ data points $\{\x_i\}$ 
from a $d$-dimensional zero-mean  Gaussian, with one component of the diagonal covariance set to 6 and all the rest to 1.
Labels $\{y_i\}$ are drawn from $y_{i}\sim \textrm{Bern}(\sigma(\w \cdot \x_{i}))$, 
where $\sigma(x)= 1/(1+e^x)$. In the regime $d\ll N$ the Laplace approximation holds fairly well, providing another good comparison method.
Figure~\ref{figNLL2SGLD} shows results for $N=1000,d=20,k=3,n=100$.

We run comparisons against the biased stochastic samplers Stochastic Gradient Langevin Dynamics (SGLD) \cite{welling2011bayesian} and multivariate Stochastic Gradient Nose-Hoover Thermostat (mSGNHT) \cite{li2015high} with fixed step sizes.  As noted above, choosing optimal step sizes for these samplers is challenging.  To allow SGLD and mSGNHT to perform best, we performed a scan to find the largest (fastest-mixing) step size that did not lead to overly large bias compared to the Laplace approximation.  (Note, importantly, that this scan is expensive and is not possible in high-dimensional examples where the Laplace approximation does not hold - precisely the cases where MCMC methods are most valuable.)
See Supp.\ Material E for  details of this  scan, which led to  an optimal step size of $0.1$ for SGLD.  Larger step sizes led to visible biases in the samples (not shown); we also show the results with step size $0.01$ for comparison to note that the results do depend sensitively on this parameter.

We also compare against ZZ-SS.  Instead of Local BPS, we ran comparisons against an unbiased method we call lipSBPS (short for Lipshitz BPS), 
where the velocity bounces occur as first arrival events in a Poisson process with noisy intensity 
$[\vv \cdot \nabla \tilde{U}(\w)]_+$ built from a noisy gradient (\ref{tU}) of minimal size $n=1$, and simulated with thinning  using an {\it exact} upper bound derived in Supp.\ Material~F. 
One can verify that the resulting stochastic process is identical to that of Local BPS. 
Our bound is higher than that used in~\cite{bouchard2015bouncy} by up to a factor of~2, which results in up to twice as many bounce proposals.
On the other hand, our bound can be computed in $O(N)$ time, does not require non-negative covariates, and can be used also for $n>1$.
Again, we note that this lipSBPS method, like Local BPS and ZZ-SS, are not generally applicable because the derived bounds only apply in special cases.

The results of Figure~\ref{figNLL2SGLD} show that SBPS outperforms the optimally tuned SGLD and mSGNHT,
and converges orders of magnitude faster than lipSBPS and ZZ-SS. While the latter two methods are unbiased, 
our results suggest that the small bias introduced by SBPS is worth the massive reduction in variance. 

In Supp.\ Material F we explore the effects of the hyperparameters: $k$, $n$, and  $\vv$ refresh rate $\lambda_r$. 
The conclusion is that in this logistic example no manual hyperparameter tuning was required (in stark contrast to the careful step size tuning required for SGLD): the bias-controlling constant $k$ can be set in the range $k \in [3,5]$ (consistent with the tails of the Gaussian in the linear regression model) and the mini-batch size $n$ should be small, but large enough for the  CLT to justify the noise term in (\ref{noisyG}); $n=100$ worked well, but the results were not sensitively dependent on $n$.  For small values of $n$ the mini-batch variability provided sufficient velocity randomness that no additional velocity refreshes were necessary, so we did not have to tune $\lambda_r$ either.

The comparison to pSBPS shows an improvement in the rate of convergence to the posterior mode.
The MAP estimator $\hat{\w}$ was calculated using SAG~\cite{roux2012stochastic}, and the Hessian was computed exactly.

\subsection{Continuous Trajectory Sampling}
\label{continuous}
\begin{figure}[t!]
  \centering
 \includegraphics[width=0.99\columnwidth]{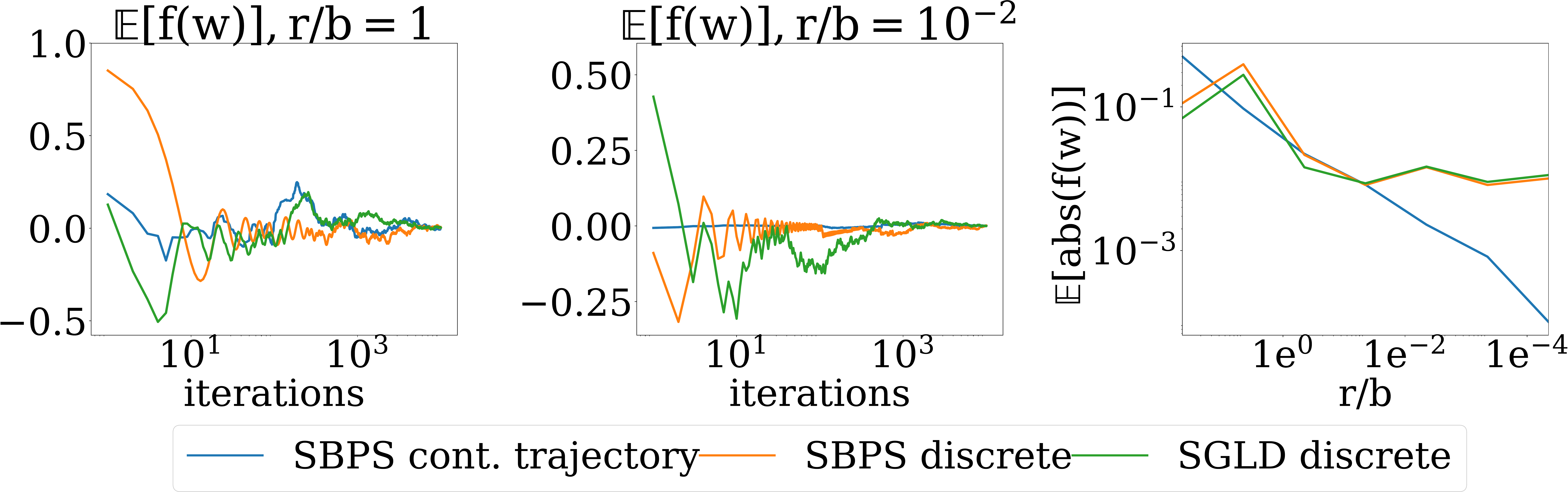}
     \caption{Estimated mean of $f(w)=\sin((w-\hat{w}/r)$, under continuous and discrete samples,
     with different ratios $r/b,$ where $b \approx 2 \times 10^{-2}$ is the average linear trajectory length. 
     The posterior distribution and settings are  as in Figure~\ref{figNLL2SGLD}.
     Assuming  the Laplace approximation holds, the expectation of $f$ is $0$. 
     \textit{Left:} For  $r/b=1$ there is little difference between continuous or discrete samples. 
     \textit{Center:} For $r/b=10^{-2}$ the continuous mean converges faster than the discrete. 
     \textit{Right:} Expectation of the absolute value of the test function averaged over 5 runs of 1000 epochs, as a function of $r/b$. The advantage of the continuous expectation when this ratio is  $r/b \ll 1$ is evident. }
\label{cont_fig}
\end{figure}

\begin{figure*}[t!]
  \centering
 \fbox{\includegraphics[width=1\textwidth]{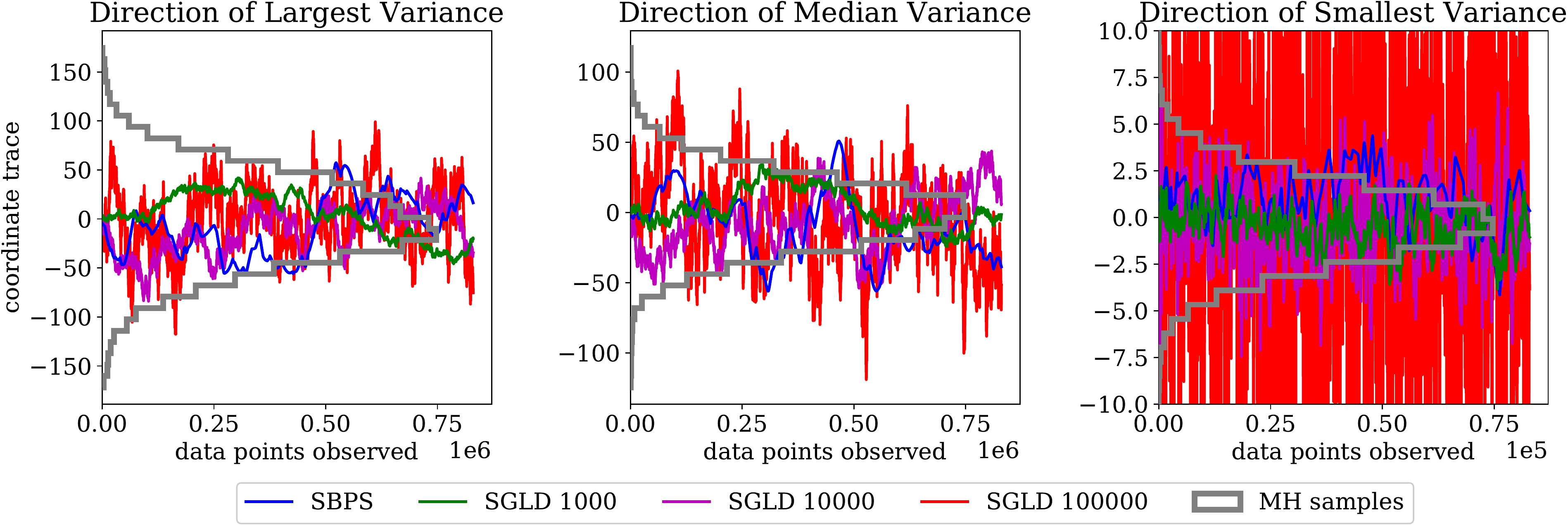}}
     \caption{Neural network posterior sampling for a single hidden layer network trained on MNIST. $d=192,N=8340,n=500$. For SBPS $k=3$. The posterior is compared to an expensive Metropolis-Hastings run. SBPS shows comparable mixing to an appropriately chosen SGLD without the need for a scan over step sizes. As can be seen, a poor choice of SGLD step size can lead to slow mixing or bias in the narrow directions of the target}
\label{figNN}
\end{figure*}
A unique feature of  BPS-like samplers is 
that their output is a continuous trajectory.
Given $\w_0$ and a set of $R$ velocities and bounce times  $\{\vv_{i},t_{i}\}$, the estimated expectation of a test function $f(\w)$ is 
\eqan
\left\langle f(\w)\right\rangle _{BPS}\equiv\frac{1}{T}\underset{i=0}{\overset{R-1}{\sum}}\underset{0}{\overset{t_{i}}{\int}} f(\w_{i}+\vv_{i}t)dt
\label{exp_bps}
\enan
where $\w_{i+1} = \w_i + \vv_i t_i$ and $T$ is the total particle travel time. For simple test functions this integral is analytic, while more generally it can be computed numerically with standard efficient one-dimensional quadrature methods. 
When  $f(\w)$ varies across a characteristic length $r$ shorter than the average trajectory length $b$ of the linear segments, 
we intuitively expect the error in the estimate~(\ref{exp_bps}) to be smaller than in estimators based on discrete samples.
Note that this advantage  tends to  diminish for higher SBPS noise, since the linear segments become shorter.

Figure~\ref{cont_fig} explores empirically this idea in a simple setting by comparing the value of the expectation of 
$f(w)=\sin( (w-\hat{w})/r)$ under the posterior distribution of the logistic example considered above. 
Here $(w,\hat{w})$ are the first coordinates of the vectors $(\w, \hat{\w})$, $\hat{\w}$ is the MAP value, and $r$ 
the characteristic length of $f$. As expected,  the error in the expectation is lower in the continuous case for $r/b<1$.

\subsection{Neural Network Posterior Sampling}
We considered  a simple model of one hidden layer followed by a softmax. 
For Bayesian approaches to neural networks see \cite{neal2012bayesian, gal2017phd}.
The likelihood  was the standard cross entropy with an additional $L_2$ regularization term 
$L=-\underset{i=1}{\overset{N}{\sum}}\log(p_i)+\frac{c}{2}\underset{j=1}{\overset{d}{\sum}}\w_{j}^{2}$ where $p_i$ is the probability of classifying the $i$th example correctly. $L$ was approximated via subsampling, and $c=0.00 1$.
This architecture was trained on the MNIST dataset. A subset of the training set was preprocessed by downsampling the images to $7\times7$, removing pixels that are 0 for all training examples and decreasing the number of digits to 4. The resulting training set size was $N=8340$. 
The resulting dimensionality of the posterior was $d=192$. Mini-batch size was $n=500$ for all methods. 
All weights were initialized at 0 and all methods were run for $10^4$ epochs. 
SBPS is compared with SGLD at different step sizes, and performance is comparable to SGLD with an appropriate step size without requiring an expensive scan over step sizes. 
Since the additional regularization term can lead to unbounded gradients of the log posterior $\nabla U(\w)$ one can no longer use the bounds derived for the Local BPS and ZZ-SS algorithms and thus they cannot be applied to this problem without further work. 
This is not the case for SBPS. 
The posterior is not Gaussian due to the likelihood terms and thus the Laplace approximation is not effective unless the posterior is dominated by the prior.

In order to assess the quality of the sampling, we compare the trajectories to a standard costly Metropolis-Hastings MCMC using a Gaussian with variance $0.2$ as the proposal distribution. This algorithm was run for $4*10^5$ epochs and the proposal acceptance rate was 0.43. 
Figure \ref{figNN} shows samples in the directions of the largest, median and smallest variance of the empirical covariance matrix of the Metropolis-Hastings samples. 


\section{Conclusions}
This paper introduced a non-reversible sampler that can be applied to big datasets by means of subsampling the data in each iteration. 
At the price of a small, controllable bias, it provides the  benefits of (i) high mixing speed associated with non-reversibility,
and (ii) continuous sample trajectories, with (iii) minimal hyperparameter tuning required, leading to state of the art performance and  making it a convenient alternative to biased, difficult-to-tune MH-based stochastic samplers.

\clearpage

\newpage
\onecolumn

\appendix
\begin{center}
{ \Large \bf Stochastic Bouncy Particle Sampler \\
\vskip .5cm 
Supplementary Material}
 \end{center}
\renewcommand{\theequation}{\thesection.\arabic{equation}}
\setcounter{equation}{0}

\vskip .5cm
\section{Proof Sketch of Invariance under Noisy Gradients}
\label{proof}

In this section we start with a simple reformulation of the proof in~\cite{bouchard2015bouncy} that the BPS Markov process leaves invariant the distribution $p(\w, \vv)= p(\w)  p(\vv)$ 
where 
\eqan 
p(\w) &\propto & e^{-U(\w)}\,, \qquad \w \in \mathbb{R}^D \,,
\label{puy2}
\\
p(\vv) &=& \mathrm{Unif}[S^{D-1}] \,,
\enan 
where $S^{D-1}$ is the $D$-dimensional one-sphere. 
This will set the stage for the noisy case considered next. For a more formal and detailed treatment of the BPS algorithm, including ergodicity, see \cite{bouchard2015bouncy}.
For simplicity, we do not include here the velocity refreshments, which do not change the proof. 

The proof sketches below are presented using a discrete-time approach followed by letting $\Delta t \rightarrow 0$. We have found this approach more accessible for  
a machine learning audience. 
After submitting a preliminary version of this work to the arXiv, the preprint~\cite{fearnhead2016piecewise} was submitted to the arXiv, which presents similar proofs of invariance by first deriving 
a general Fokker-Planck equation and then showing that the equation is satisfied both in noiseless and noisy cases.

\subsection{Exact Gradient}
To understand why the algorithm is correct, consider first the transition rule  
\eqan 
(\w, \vv)_{t + \Delta t} = 
\left\{
 \begin{array}{ll}
 (\w  + \vv \Delta t  , \vv)   &  \textrm{with probability} \,\,  1-\Dt [ G]_+ 
\\
(\w + \vv \Delta t , \vv_r) &  \textrm{with probability} \,\,  \Dt [G]_+ 
\end{array}
\right.  
\enan 
where 
\eqan 
[x]_{+} = \max(x,0) \,,
\\
G =  \vv \cdot \nabla U(\w) \,,
\enan 
and 
\eqan 
\vv_r =  \vv -2\frac{(\vv \cdot \nabla U(\w)) \nabla U(\w) }{||\nabla U(\w)||^2} \,.
\enan 
This rule acts on the probability density $p(\w, \vv)$ as,
\eqan 
p_{t+ \Delta t}(\w, \vv) 
&=&  [p_{t+ \Delta t }(\w, \vv) ]_d + [p_{t+ \Delta t }(\w, \vv) ]_r \,.
\label{pkp2}
\enan 
The two terms in  (\ref{pkp2}) correspond to the two ways  to reach  $(\w, \vv)$ at time $t+\Delta t$. 
First, we can start  at $(\w -\vv\Delta t, \vv)$ at time $t$ and move a distance $\vv\Delta t$ without bouncing. This occurs with probability $1-\Delta t [\vv \cdot \nabla U]_+$, 
so we have
\eqan 
[p_{t+ \Delta t }(\w, \vv) ]_d &=& (1-\Delta t [\vv \cdot \nabla U]_+ ) p_t(\vv) p_t(\w-\vv \Delta t)  \,,
\label{pd}
\\
 &=& (1-\Delta t [\vv \cdot \nabla U]_+ ) p_t (\vv) (p_t(\w) - \Delta t   \vv \cdot \nabla p_t (\w) + O(\Delta t^2))  \,,
\label{taylor1}
\\
&=& p_t (\vv) p_t (\w) \left[ 1 + \Delta t   \vv \cdot \nabla U   - \Delta t [\vv \cdot \nabla U]_+ \right] + O(\Delta t^2) \,,
\label{usepuy}
\enan 
where in (\ref{taylor1}) we did a Taylor expansion and in (\ref{usepuy}) we used (\ref{puy2}).

The second term in (\ref{pkp2}) corresponds to  being at $(\w -\vv_r\Delta t, \vv_r)$ at time $t$, moving $\vv_r\Delta t$ and bouncing.  This occurs with probability
$\Delta t [\vv_r \cdot \nabla U]_+ =\Delta t [-\vv \cdot \nabla U]_+$, so we have 
\eqan 
[p_{t+ \Delta t }(\w, \vv) ]_r &=& \Delta t [-\vv \cdot \nabla U]_+ p_t(\w-\vv_r \Delta t, \vv_r) \,,
\label{tty}
\\
&=& \Delta t [-\vv \cdot \nabla U]_+ p_t(\w, \vv_r) + O(\Dt^2)  \,,
\label{taylor2}
\enan 
where again we did a Taylor expansion in (\ref{tty}). Adding (\ref{usepuy}) and (\ref{taylor2}), and using 
\eqan 
[\vv \cdot \nabla U]_+ - [-\vv \cdot \nabla U]_+ = \vv \cdot \nabla U \,,
\label{sumvu}
\enan 
equation~(\ref{pkp2}) becomes
\eqan 
p_{t+ \Delta t }(\w, \vv)  &=&  p_{t }(\w, \vv)  +  O(\Delta t^2) \,,
\label{expt}
\enan 
which implies that the distribution is stationary, $\frac{d p_t (\w, \vv )}{dt} =0$. 

\subsection{Noisy Gradient}

Consider now a noisy gradient represented as
\eqan 
\nabla \tilde{U}(\w) = \nabla U(\w) + {\bf n}_\w \,, \qquad \qquad {\bf n}_\w \sim  p(  {\bf n}_\w |\w)\,, \quad     \quad {\bf n}_\w \in {\mathbb R}^D \,,
\label{unoise2}
\enan 
where we assume that $p({\bf n}_w |\w)$ has zero mean.

First note that the requirement that ${\bf n}_\w$ and ${\bf n}_\w'$ are conditionally independent given $\w$ and $\w'$, with $\w \neq \w'$, is needed to preserve under the noise the Markov property of the sampler, 
which requires the bounce point process intensity to depend only on $\w$, and not the past history of the $\w$ trajectory.

Next we  decompose the random vector ${\bf n}_\w $ into two orthogonal components,
\eqan 
{\bf n}_\w = y \vv + {\bf n}_{\vv} \,,
\enan 
with $y = \vv \cdot {\bf n}_\w$, and ${\bf n}_{\vv} \cdot \vv =0$. This induces a corresponding decomposition in the probability density as  
\eqan 
d {\bf n}_\w p({\bf n}_\w|\w) = dy d{\bf n}_{\vv} p(y|\w) p({\bf n}_{\vv} | y, \w, \vv) \,,  
\label{decomp}
\enan 
and note that from the assumption that $p({\bf n}_\w |\w)$ has zero mean it follows that $p(y|\w)$ has zero mean. 
The noisy projected gradient  becomes 
\eqan
\vv \cdot \nabla U(\w)   +y \,, \qquad  y \sim p(y|\w) \,.
\label{gnoise}
\enan 
To study the invariance of $p(\w, \vv) $ under the noisy BPS, let us consider again the  decomposition~(\ref{pkp2}) into straight and bounced infinitesimal trajectories.
The probability  that the particle is at $(\w -\vv\Delta t, \vv)$ at time $t$ and moves a distance $\vv\Delta t$ without 
bouncing is the average of $1-\Dt [\vv \cdot \nabla U(\w)   +y]_+$ over all the possible realizations of $  y$, and is therefore given by 
\eqan 
1-\Dt    P_{\vv} &\equiv& 1-\Dt \int_{-\infty}^{+\infty} \!\!\!  [ \vv \cdot \nabla U(\w)   +  y]_{+}   p(y|\w) dy \,,
\\
&=& 1- \Dt \int_{-\vv \cdot \nabla U(\w)}^{+\infty} \!\!\!  \,   ( \vv \cdot \nabla U(\w)   +  y)  p(y|\w) dy \,,
\label{int1}
\enan 
where the above expression defines $P_{\vv}$. The first term of (\ref{pkp2}) is therefore
\eqan 
[p_{t+ \Delta t }(\w, \vv) ]_d &=&  ( 1-\Dt    P_{\vv}) p(\w - \vv \Dt, \vv) \,,
\\
&=& p_{t}(\w, \vv)  - \Delta t   \vv \cdot \nabla p_t (\w) p_t(\vv) - \Dt P_{\vv} p_t (\w) p_t(\vv) + O(\Dt^2)   \,,
\nonumber 
\\
&=& p_{t}(\w) p_{t}(\vv) [ 1  + \Delta t   \vv \cdot \nabla U(\w)      - \Dt P_{\vv}  ] + O(\Dt^2)   \,,
\label{pd2}
\enan 
similarly to (\ref{pd})-(\ref{usepuy}).

The second term in (\ref{pkp2}) now has contributions from all those values 
$(\w -\tilde{\vv}_r\Delta t, \tilde{\vv}_r)$ at time $t$, such that a reflection of $\tilde{\vv}_r$ with respect to  a noisy $\nabla \tilde{U}(\w)$ gives $\vv$. 
Such a  $\tilde{\vv}_r$ exists for every value of the noise vector $ {\bf n}_w$, and is given by
\eqan 
\tilde{\vv}_r   =  \vv -2\frac{( \vv \cdot \nabla \tilde{U}(\w)) \nabla \tilde{U} (\w) }{||\nabla \tilde{U} (\w)||^2} \,,
\label{vvr2}
\enan 
Therefore the  second term in (\ref{pkp2}) contains contributions from all the possible  realizations of ${\bf n}_\w$ and is 
\eqan 
[p_{t+ \Delta t }(\w, \vv) ]_r  
&=& \Dt \int_{{\mathbb R}^D} d {\bf n}_\w  [ \tilde{\vv}_r \cdot \nabla \tilde{U}(\w)]_{+}  p({\bf n}_\w| \w ) p_t(\w- \tilde{\vv}_r \Delta t, \tilde{\vv}_r) \,,
\\
&=& \Dt p_t(\w, \tilde{\vv}_r) \! \int_{-\infty}^{+\infty} \!\!\!\!\!\! dy \, p(y|\w) [ - \vv \cdot \nabla U(\w)   -  y]_{+} \,,
\times \int \!\!\! d{\bf n}_{\vv} p({\bf n}_{\vv} | y, \w, \vv)    + \, O(\Delta t^2) \,,
\nn
\\
&=& \Dt P_{\vv_r} p_t(\w, \tilde{\vv}_r) + O(\Delta t^2) \,,
\label{pr2}
\enan 
where we used $\tilde{\vv}_r \cdot \nabla \tilde{U}(\w) = - \vv \cdot \nabla U(\w)   -  y $, the measure decomposition (\ref{decomp}),  $\int \!\!\! d{\bf n}_{\vv} p({\bf n}_{\vv} | y, \w, \vv)=1$ 
and defined 
\eqan 
P_{\vv_r} &=& \int^{ - \vv \cdot \nabla U(\w)  }_{-\infty} \!\!\! d  y \,   (-  \vv \cdot \nabla U(\w)   -   y)  p(y|\w) \,.
\label{int2}
\enan 
Adding now (\ref{pd2}) and (\ref{pr2}), using $p(\tilde{\vv}_r)= p(\vv)$ (since $p(\vv)$ is uniform) and 
\eqan 
P_{\vv} - P_{\vv_r} =  \vv \cdot \nabla U(\w) \,,
\enan 
which follows from (\ref{int1}) and (\ref{int2}), and the fact that $p(y|\w)$ has zero mean,
we get again the stationarity condition
\eqan 
p_{t+ \Delta t }(\w, \vv)  &=&  p_{t }(\w, \vv)  +  O(\Delta t^2) \,.
\label{expt2}
\enan

\section{Biased Approximation}
\subsection{Biased bouncing rate}
In the noiseless case, the velocity bounce is an event in a Poisson process with intensity
$\lambda(\w) = [\vv \cdot \nabla U(\w)]_{+}$
while in the noisy case, the average Poisson intensity is
$\lambda_n(\w) = E_y [\lambda_n(\w,y) ]$
where 
\eqan
\lambda_n(\w,y) = [\vv \cdot \nabla U(\w) + y]_{+} \,.
\enan 
When a thinning upper bound for $[\vv \cdot \nabla U + y]_{+}$ is unknown and the distribution of $y$ is Gaussian with predicted variance~$\rho^2$,
our algorithm makes a bounce proposal from a Poisson process with intensity 
\eqan 
\lambda_{\rho}(\w) = \hat{G} + k \rho(\w)\,,
\enan
where $\hat{G}$ is our estimate of $\vv \cdot \nabla U(\w)$. 
At the proposed bounce point $\w$, we evaluate $\lambda_n(\w,y)$, and  accept with probability $\min(\lambda_n(\w,y)/\lambda_{\rho}(\w), 1)$.
The evaluation of $\lambda_n(\w,y)$ also provides an estimate $\sigma^2(\w)$ of the variance of $y$.
 Assuming $y$ is Gaussian, the probability of the bound violation event  $1 < \lambda_n/  \lambda_{\rho}$,  is 
 \eqan
 q(\w) = 1-\Phi((\lambda_{\rho}(\w) - \vv \cdot \nabla U(\w) )/ \sigma(\w) ) \,,
 \enan 
where $\Phi$ is the standard normal CDF. For a given $y$, the intensity is therefore,
\eqan 
\lambda_b (\w,y) &=& I_{ [ \frac{\lambda_n}{  \lambda_{\rho}} < 1 ]} \lambda_n(\w,y) + I_{ [\frac{\lambda_n}{  \lambda_{\rho}} > 1 ]} \lambda_{\rho}(\w)
\enan 
where $I_{[\cdot]}$ is the indicator function.
Averaging over $y$ we get
\eqan 
\lambda_b (\w) &=& E_y[\lambda_b (\w,y)]
\\
&=& (1-q (\w)) E_{ \lambda_n \leq \lambda_{\rho} } [\lambda_n(\w,y)] + q(\w) \lambda_{\rho}(\w)
\label{lambda_b}
\enan 
If the probability of bound violation has a universal upper bound $ q(\w) < q, \forall \w$,  we assume 
\eqan 
|\lambda_b (\w)  - \lambda_n(\w)| \leq K_q = Cq + O(q^2)
\label{kbound}
\enan 
where $C$ is a constant.


\subsection{Preliminaries}
We are interested bounding the distance between 
the equilibrium distribution of the biased, noisy BPS process with mean intensity $\lambda_b(\w)$,
and the exact, noisy process with mean intensity $\lambda_n(\w)$. We start with some preliminary results.

\subsection*{Wasserstein Distance  and Kantorovich Duality}
We will consider the Wasserstein distance, defined as 
\eqan 
d_{\cal W}(p_1, p_2) = \sup_{f \in C_L} |E_{p_1}[f] - E_{p_2}[f]|  \,,
\enan 
where $C_L$ is the set of 1-Lipshitz  continuous functions,
\eqan 
C_L = \{f:\mathbb{R}^d \rightarrow \mathbb{R} : |f(y)- f(x) |  \leq |y-x| \}  \,.
\label{cl}
\enan 
Given random variables $\z_1 \sim p_1, \, \z_2 \sim p_2$, a coupling is a joint distribution $(\z_1, \z_2) \sim p_{12}$
with  marginals $p_1$ and $p_2$.
The Kantorovich duality~\cite{villani2008optimal} asserts that 
\eqan
d_{\cal W}(p_1, p_2) = \inf_{p_{12}} E_{p_{12}} [|\z_1-\z_2|]  \,.
\label{kantorovich}
\enan 

\subsection*{Generators}
To simplify the notation, let us define $\z= (\w,\vv)$, $\z_r= (\w,\tilde{\vv}_r)$. 
The infinitesimal generator of a stochastic process is defined as 
\eqan 
{ \cal L} f(\z) = \lim_{\delta t \rightarrow 0} \frac{ E[ f(\z_{t+ \delta t} )  |\z_t=\z] - f(\z) }{\delta t} \,,
\enan 
and note that it satisfies
\eqan 
E[{\cal L} f] &=& \lim_{\delta t \rightarrow 0} \frac{ \int d\z_{t+ \delta t} d\z p(\z_{t+ \delta t}  |\z) p(\z)  f(\z_{t+ \delta t} )   - E[f(\z)] }{\delta t} \,,
\\
&=& \lim_{\delta t \rightarrow 0} \frac{ \int d\z_{t+ \delta t} p(\z_{t+ \delta t})  f(\z_{t+ \delta t} )   - E[f(\z)] }{\delta t} \,,
\\
&=& 0,
\label{elfz}
\enan 
where the expectation  is with respect to the distribution $p(\z)$ invariant under the stochastic process, and we used 
$\int d\z p(\z_{t+ \delta t}  |\z) p(\z) = p(\z_{t+ \delta t}) $.
In our case, the generator of a BPS process with  intensity $\lambda_n(\w,y)$ is~\cite{davis1984piecewise, fearnhead2016piecewise} 
\eqan
{\cal L}_{\lambda_n} f(\z) = \vv \cdot \nabla_{\w} f(\z) +  E_y [\lambda_n(\w,y)  (f(\z_r) -  f(\z) )]
\label{generator}
\enan 
and similarly for  $\lambda_b(\w)$.

Let us define 
\eqan 
f_{\lambda}(\z,t) = E_{\lambda} [ f(\z_{t}) |\z_0=\z] \,,
\label{fzt}
\enan 
where the expectation is with respect to the distribution of the stochastic process with intensity $\lambda$ at time $t$ and with a given initial condition.
This expression  satisfies the backward Kolmogorov equation 
\eqan
\frac{\partial f_{\lambda}(\z, t) }{\partial t}&=&  {\cal L}_{\lambda} f_{\lambda}(\z,t) \,,
\label{back_kol}
\enan
and also~\cite{jacod1987limit}
\eqan 
\lim_{t \rightarrow \infty} f_{\lambda}(\z,t) = E_{\lambda}[f] \,,
\label{lim_infty}
\enan 
where the expectation $E_{\lambda}[\cdot]$ is with respect to the distribution invariant under the 
stochastic process with intensity $\lambda$.

\subsection*{Ergodicity}
We assume that the random process defined by SBPS is polynomial 
ergodic (although see the recent~\cite{deligiannidis2017exponential}). In particular, we assume that two distributions
started at reflected velocities $p_{\lambda_n,t,\z} = p_{\lambda_n}(\z_t|\z_0 = \z), \, p_{\lambda_n,t,\z_r} = p_{\lambda_n}(\z_t|\z_0 = \z_r)$ converge as
\eqan 
d_{\cal W}(p_{\lambda_n,t,\z}, p_{\lambda_n,t,\z_r}) \leq \frac{C_A}{(\alpha + t)^{\beta}}
\label{ergo}
\enan 
where $\alpha, \beta, C_A$ are 
constants.\footnote{This assumed property follows usually from the existence of small sets (Lemma 3 in \cite{bouchard2015bouncy}) 
along with an appropriate Lyapunov function~\cite{roberts2004general}.}

\subsection*{Poisson Equation}
Given a  function $f(\z)$, we will consider below the Poisson equation
\eqan 
{\cal L}_{\lambda} u_f(\z) = f(\z) -E_{\lambda}[f] \,.
\label{poisson}
\enan 
We assume the existence of the solution
\eqan 
u_f(\z) = \int_0^{\infty} ds (  E_{\lambda}[f] - f_{\lambda}(\z,s) ) \,,
\enan 
where $f_{\lambda}(\z,s)$ was defined in (\ref{fzt}). 
The fact that this expression solves (\ref{poisson}) can 
be easily verified using (\ref{back_kol}), (\ref{lim_infty})   and $f_{\lambda}(\z,0) = f(\z)$. 
For $f \in C_L$ (see (\ref{cl}) ), this solution satisfies
\eqan 
|u_f(\z)- u_f(\z_r)|  &=& \left| \int_0^{\infty} ds ( f_{\lambda}(\z,s) - f_{\lambda}(\z_r,s) ) \right| 
\\
&\leq& \int_0^{\infty} ds E_{\lambda}[ \z_s - \z_{r,s}| ]      \,, \quad \textrm{using the Lipshitz property }
\\
&\leq& \int_0^{\infty} ds \, d_{\cal W}(p_{\lambda,s,\z}, p_{\lambda,s,\z_r})   \,, \quad \textrm{using~(\ref{kantorovich}) }
\\
&\leq& \frac{C_A}{(\beta+1)\alpha^{\beta+1}}     \,, \quad \textrm{using the ergodicity assumption (\ref{ergo})} \,.
\label{uud}
\enan

\subsection{ Distance Bound from Stein's Method}
We now prove a bound on the distance between the exact and biased distributions, using Stein's method~\cite{barbour1990stein, ross2011fundamentals, stein1972bound},
which was recently used for the related Zig-Zag process~\cite{huggins2016quantifying}.

\eqan 
d_{\cal W}(p_{\lambda_n}, p_{\lambda_b}) &=& \sup_{f \in C_L} |E_{{\lambda_n}}[f] - E_{{\lambda_b}}[f]|  \,,
\label{dfirst}
\\
&=& \sup_{f \in C_L} |E_{{\lambda_n}}[{\cal L}_{\lambda_b} u_f]|  \,, \quad \textrm{using (\ref{poisson})}
\\
&=& \sup_{f \in C_L} |E_{{\lambda_n}}[{\cal L}_{\lambda_b} u_f] - E_{{\lambda_n}}[{\cal L}_{\lambda_n} u_f]|  \,, \quad \textrm{using (\ref{elfz})}
\\
& \leq & \sup_{f \in C_L} E_{{\lambda_n}}[ |({\cal L}_{\lambda_b}-{\cal L}_{\lambda_n})  u_f |]   \,.
\enan 
Note that this last expression involves an integral over just one distribution, unlike the first expression (\ref{dfirst}).
Inside the expectation we have, using (\ref{generator}), 
\eqan 
({\cal L}_{\lambda_n}-{\cal L}_{\lambda_b})  u_f(\z)  &\leq& |\lambda_n(\w) - \lambda_b(\w) | E_y [ |u_f(\z_r) - u_f(\z) |] \,,
\enan 
where the expectation over $y$ is because  $\z_r = (\w, \vv_r)$ depends on the noise $y$ (see (\ref{vvr2})).
Using (\ref{kbound}) and (\ref{uud}), we get finally
\eqan 
d_{\cal W}(p_{\lambda_n}, p_{\lambda_b}) \leq  \frac{K_q C_A}{(\beta+1)\alpha^{\beta+1}} \,.
\enan
Interestingly, this bound depends on the mixing speed of the  process generated by ${\cal L}_{\lambda_n}$ (see (\ref{ergo})),
even though the distance is between two {\it equilibrium}  distributions.

\newpage

\section{SBPS algorithm}
\label{alg2box}
Algorithm~\ref{sbpsalgo} provides a description of the SBPS algorithm with a linear regression based thinning proposal intensity. We have omitted velocity refreshments for the sake of clarity. $\Delta t$ in the code below is the resolution of the piecewise linear proposal intensity, which should be smaller than the typical time between bounces. In all experiments a value of $\Delta t = .01$ was used.

\begin{algorithm}[H]
  \begin{tcolorbox} 
  \caption{Stochastic Bouncy Particle Sampler}
  \label{sbpsalgo}  
      \begin{algorithmic}
      	  \STATE \textbf{SBPS:}
	  \STATE Initialize particle position $ \w \in \mathbb{R}^D$, velocity $\vv \in S^{D-1}$, $t \leftarrow 0$, regression coefficients $\hat{\beta}_0,\hat{\beta}_1,\rho(t)$
	  \WHILE{$\text{desired}$}
	  	  \STATE $t,\lambda(t) =$ Sample\_Proposal\_Time($\hat{\beta}_0,\hat{\beta}_1,\rho(t)$)
		  \STATE $\w  \leftarrow \w + \vv*t$
		  \STATE Store $\w,t$
	  	  \STATE Observe $\nabla \tilde{U}(\w),\textrm{Var}  \left[ \vv \cdot  \nabla \log p(x_{r_i}|\w)  \right ]$
		  \STATE (optional: Update preconditioner and apply it to gradient - see \ref{prec_BPS})
		  \STATE Calculate $\tilde{G}(t),c(t)$ 
		  \STATE $\vv =$ Accept/Reject\_Proposal($\tilde{G}(t),\lambda(t),\vv$)
		  \STATE $\hat{\beta}_0,\hat{\beta}_1,\rho(t) = $ Update\_Local\_Regression\_Coefficients($\tilde{G}(t),c(t),t$)
	  \ENDWHILE
	  \STATE  Return piecewise linear trajectory of $\w$
	  \STATE
	   \STATE \textbf{Sample\_Proposal\_Time($\hat{\beta}_0,\hat{\beta}_1,\rho(t)$):}
	  \STATE $t_{next\_proposal} \leftarrow 0$
	  \STATE Initialize set of interpolation points $p = \{[\hat{\beta}_{1}t_{next\_proposal}+\hat{\beta}_{0}+k \rho(t_{next\_proposal})]_+\}$
	  \STATE Initialize piecewise linear proposal intensity $\lambda(t) = Inter(p)$*
	  \STATE Sample $u \sim \mathrm{Unif}[0,1]$
	   \WHILE{$-log(u) > \int_{0}^{t_{next\_proposal}}{\lambda(t)dt}$}
	   \STATE $t_{next\_proposal} \leftarrow t_{next\_proposal} + min(\Delta t,-log(u) - \int_{0}^{t_{next\_proposal}}{\lambda(t)dt})$
	   \STATE $p \leftarrow p \cup [\hat{\beta}_{1}t_{next\_proposal}+\hat{\beta}_{0}+k \rho(t_{next\_proposal})]_+$
	   \STATE $\lambda = Inter(p)$
	  \ENDWHILE 
	  \STATE Return $t_{next\_proposal},\lambda(t_{next\_proposal})$
	  \STATE * $Inter(p)$ is a linear interpolation of the points in $p$ and their respective times since the last proposal

%
  	  \STATE \textbf{Accept/Reject\_Proposal($\tilde{G}(t),\lambda(t),\vv$):}
	  \STATE Draw $u \sim \mathrm{Unif}[0,1]$
  	   \IF {$u > \tilde{G}(\w)/\lambda(t)$}
		  \STATE Proposed bounce time accepted:
		  \STATE Initialize $\{\tilde{G}(t_i),c(t_i)\}$ and regression coefficients $\hat{\beta}_0,\hat{\beta}_1,\rho(t)$ using $\tilde{G}(t),c(t)$
		  \STATE Return $\vv -2\frac{(\vv \cdot \nabla \tilde{U}(\w)) \nabla \tilde{U}(\w) }{||\nabla \tilde{U}(\w)||^2}$ 
  	   \ELSE
	        \STATE Proposed bounce time rejected, maintain current trajectory:
	   	\STATE Return \vv
 	   \ENDIF
	   \STATE
	  \STATE \textbf{Update\_Local\_Regression\_Coefficients($\tilde{G}(t),c(t),t$):}
	  \STATE Add $\tilde{G}(t),c(t)$ to $\{\tilde{G}(t_i),c(t_i)\}$
	  \STATE (optional: Perform hyperparameter learning step on regression priors)
	  \STATE Update regression coefficients $\hat{\beta}_0,\hat{\beta}_1,\rho(t')$ using standard Bayesian regression formula
	  \STATE (optional: If $\hat{\beta}_1 < 0$ set $\hat{\beta}_1$ to non-negative value, update $\hat{\beta}_0$ accordingly)
	  \STATE Return $\hat{\beta}_0,\hat{\beta}_1,\rho(t)$
      \end{algorithmic}
  \end{tcolorbox}
\end{algorithm}

\section{A Highly non-Gaussian Example}
\label{ring}

\begin{figure*}[t!] 
  \centering
 \fbox{\includegraphics[width=1\textwidth]{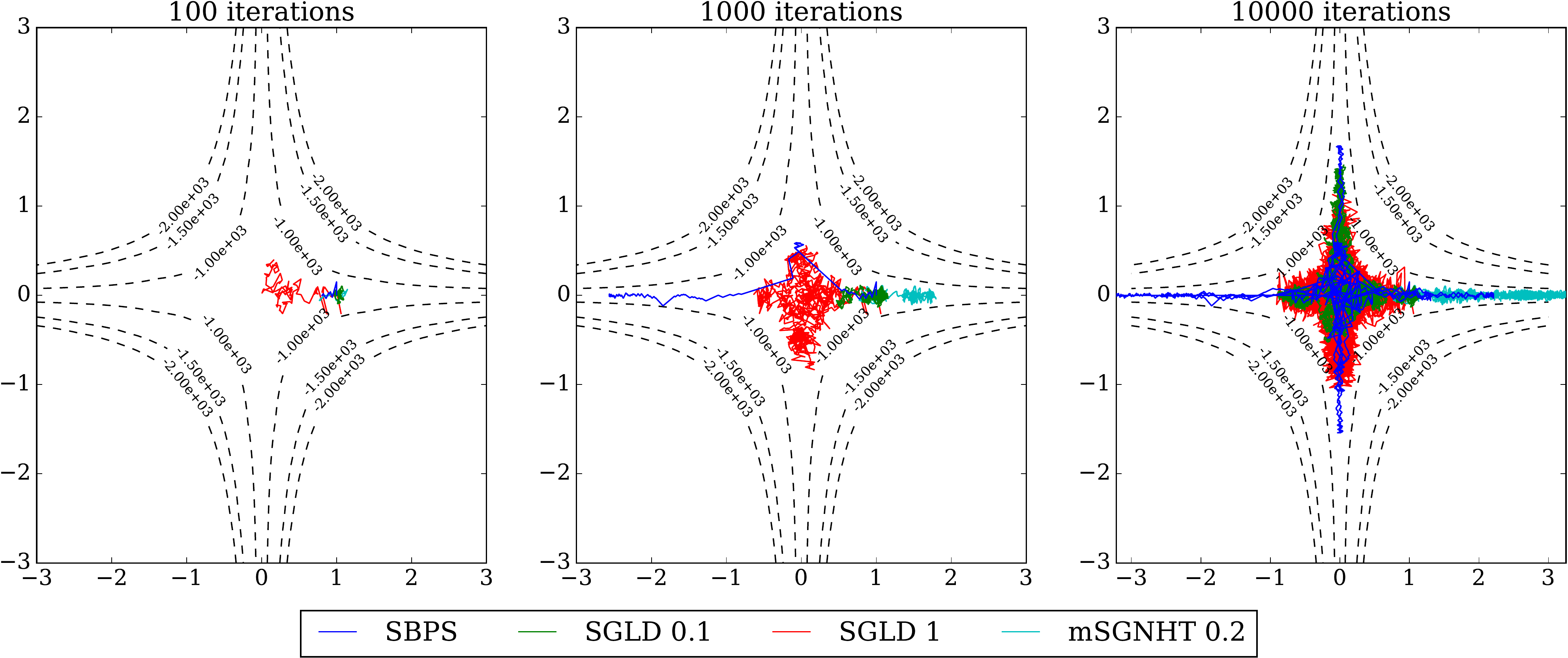}}
     \caption{Sample traces of SBPS, SGLD and mSGNHT sampling a highly non-Gaussian posterior. SBPS appears to explore the posterior more fully and avoids regions of low density as opposed to the large step size SGLD, leading to less bias. 
         }
\label{figKL}
\end{figure*}

We explored a case of sampling from a Bayesian posterior where the Laplace approximations is not accurate. Figure~\ref{figKL} shows results for a highly non-log-concave 2D hyperboloid posterior, with data generated according to $y_{i}\sim\mathcal{N}(w_{0}^*w_{1}^*,\sigma)$. After introducing a weak Gaussian prior, the resulting log posterior takes the form 
\eqan
L(w_0, w_1|\{y_{i}\},\sigma)=\underset{i=1}{\overset{N}{\sum}}-\frac{(y_{i}-w_{0}w_{1})^{2}}{2\sigma^{2}}-\frac{c}{2}||w||_{2}^{2} \,.
\enan 
This posterior was approximated by observing mini-batches of data as in the previous examples. The scaling symmetry that is manifest in the invariance of the likelihood with respect to $w_{0},w_{1}\rightarrow\lambda w_{0},\frac{w_{1}}{\lambda}$ leads to the highly non-Gaussian hyperboloid form. Similar symmetries are encountered in posteriors of deep neural networks with ReLU activation \cite{dinh2017sharp}. The parameters used were $N=1000,n=100,k=3,c=.0001,w_{0}^*=w_{1}^*=0,\sigma=1$. $\sigma$ was not learned. 

Figure~\ref{figKL} shows comparisons with SGLD and mSGNHT, while  Local BPS and SS-ZZ 
cannot be applied since there seems to be no simple exact upper bound for thinning in this case. 
Note that for the step sizes shown, both SGLD and mSGNHT deviate into low density regions while not mixing as well as SBPS. The smaller SGLD step size used does not deviate as much but exhibits even slower mixing. 

\section{Sampling from Multimodal Targets}
Our simple  linear model for $G(t)$ (the projected gradient of the log posterior),
appears to be sufficiently accurate even when the Laplace approximation is violated (as shown in the previous examples),
but in some highly multimodal cases we have found this approximation to be insufficient.

In this section we present a slight modification of SBPS to sample from such targets as well. 
The potential troubles arise because in a multimodal target one may encounter situations where the measured 
$G(t)$ drop quickly between successive observations, leading to strong negative regression slopes. 
In our regression model, we get an interpolation (cf. equation~(13)) leading to an upper bound of the form
\eqan
\hat{\beta}_{1}t+\hat{\beta}_{0}+k \rho(t) 
\label{up_bound}
\enan
with $\hat{\beta}_{1} < 0$. Since the leading order $t$ dependence in $\rho(t)$ is linear, if $k$ is too small the linear term in (\ref{up_bound}) may be negative. 
This will lead the sampler to propose long times between samples and thus enter low target density regions of the space. 

In such cases, we propose to make additional auxiliary observations at times $\{t_{aux}\}$ along the current linear trajectory of the particle and update the linear bound accordingly before making the next proposal. On a large enough scale this procedure will make auxiliary observations $\tilde{G}>0$ leading to a positive slope in (\ref{up_bound}). This in turn will prevent the particle from entering low target density regions. 

Note that these auxiliary observations can be performed with the same minibatch of data from the last bounce proposal.
In principle, such strong negative slopes can occur even for a unimodal target if the subsampling noise is highly non-Gaussian, 
and this mechanism can also be used in those situations.

\begin{figure*}[t!] 
  \centering
 \fbox{\includegraphics[width=.6\textwidth]{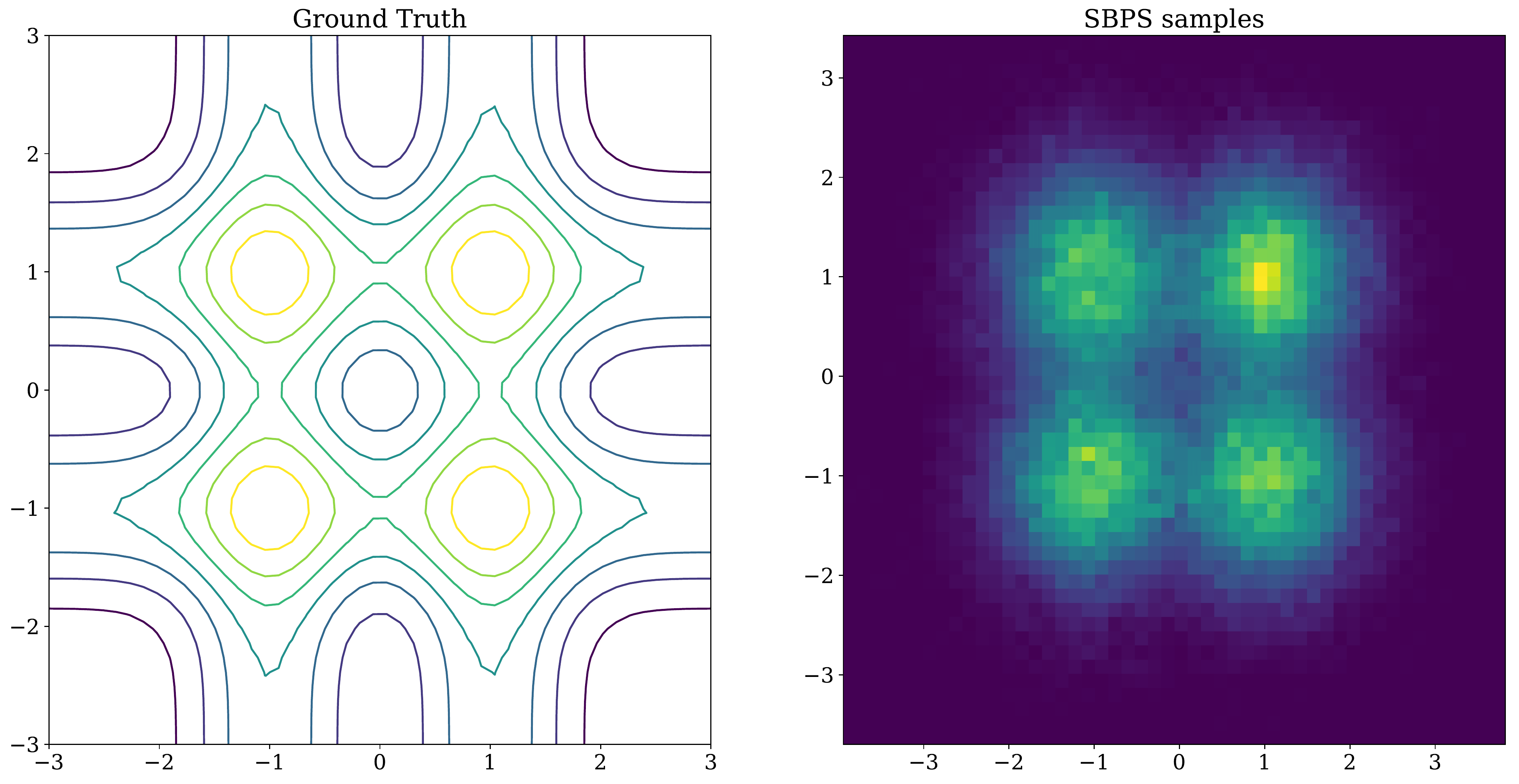}}
     \caption{SBPS sampling from a highly multimodal target. As can be seen from the sample histogram on the right, SBPS manages to accurately capture the multimodal target. The results are from 1000 epochs of sampling from a dataset of size $N=1000$}
\label{figmultimodal}
\end{figure*}

We illustrate this mechanism in a simple distribution defined as 
\eqan
L(w)=\sum_{i=1}^N \underset{k=1}{\overset{D}{\sum}}L_{i}(w_{k}) + \textrm{const.}
\enan 
where $w \in \mathbb{R}^D$ and 
\eqan
L_{i}(w_{k}) = \log\left[ e^{- \frac{(w_{k}-1-\mu_{k}^{i})^2} {2\sigma_L^2}} 
+
e^{- \frac{(w_{k}+1-\mu_{D+k}^{i})^2} {2\sigma_L^2}}
\right]
\enan 
and each ${\mu^i_k}$ is drawn from $\mathcal{N}(0,\sigma_\mu)$. 
This is a highly multimodal toy distribution. Although it does not come from a posterior distribution, it allows us to illustrate the proposed mechanism in a clean setting. 
Figure \ref{figmultimodal} shows  results for $D=2$, $N=1000,\sigma_L=.25,\sigma_\mu=.01$ and mini-batch size  $n=10$. 
We used $\{ t_{aux} \}=\{10p\overline{t},p \in \mathbb{N} \}$ where $\overline{t}$ is the mean proposal time of all past proposals during the sampling process. 
While one can add multiple auxiliary points in this way, in practice we have found that  one auxiliary  point ($p=1$) is sufficient. 
Figure \ref{figmultimodal} shows that SBPS is able to correctly sample from the target while avoiding the issues posed by the multimodality of the distribution. 
We note that this modified mechanism does not affect the rest of the examples presented in this paper.

%
%

\section{SGLD Step Size Scan}
\label{stepscan}

\begin{figure}[t!] 
  \centering
  \includegraphics[width=.45\textwidth]{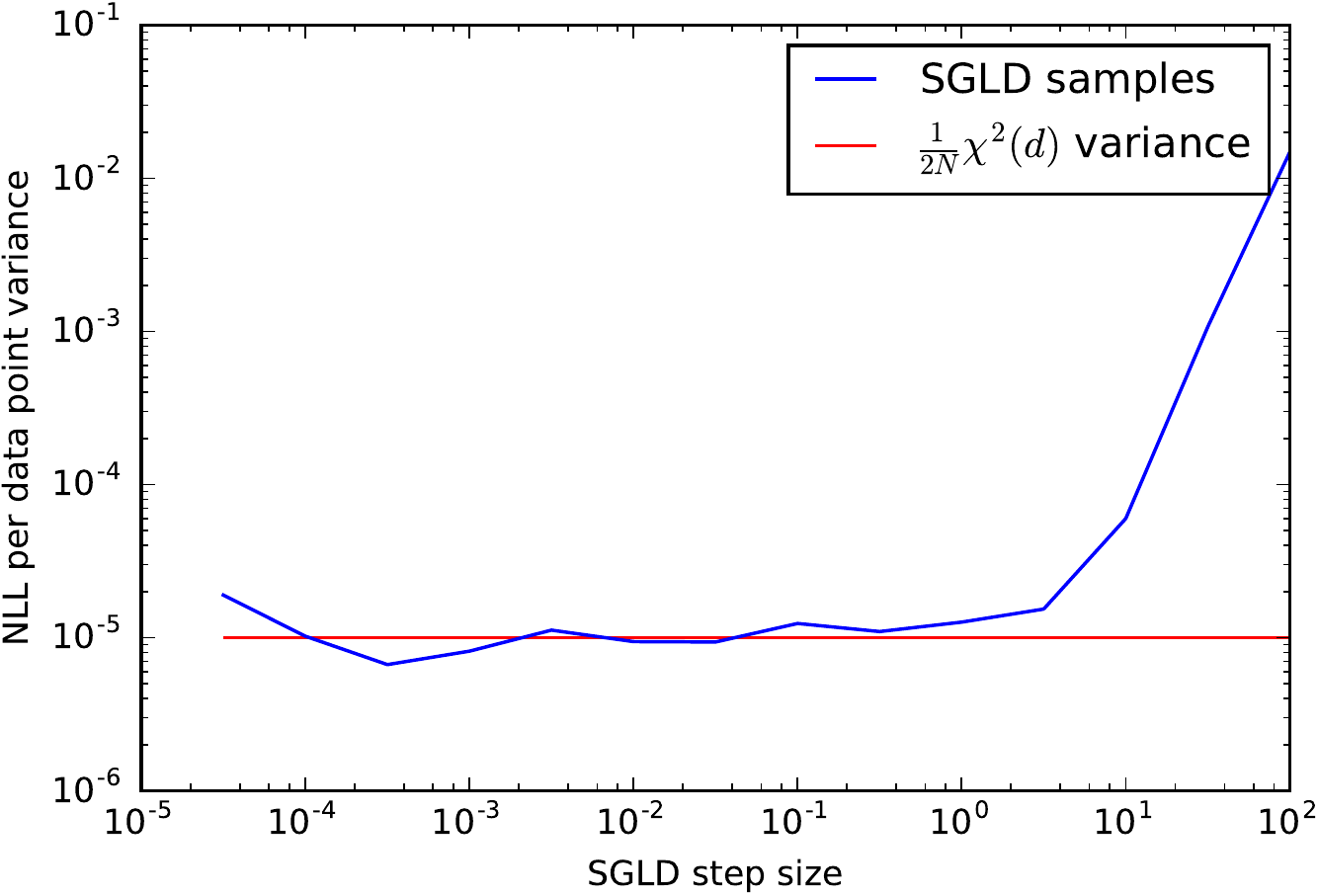}
     \caption{Per-data-point variance of the NLL in the logistic regression example of Section~7.1, using SGLD samples with step sizes of $10^{-i/2},i=0...9$. The samplers were 
      initialized at the MAP. 
      We select the biggest step size whose  empirical variance is below that from the Laplace approximation, $\frac{d}{2N^2}$.}
\label{figstep2}
\end{figure}

In the logistic regression example of Section~6.1, we compare SBPS with Stochastic Gradient Langevin Dynamics (SGLD)~\cite{welling2011bayesian} with fixed step size.
A natural question is how to choose an appropriate step size 
that ensures the fastest possible mixing  without introducing an unacceptable amount of bias. 
Our criterion was to pick the biggest possible (i.e., fastest-mixing) step size such that the resulting variance of the 
per-data-point  Negative Log Likelihood (NLL) coincides with that of the Laplace approximation. 
The latter gives a per-data-point NLL distribution of $\frac{1}{2N}\chi^{2}(d)+NLL_{\hat{\w}}/N $ where $\hat{\w}$ is the MAP estimator~\cite{bickel2015mathematical}.
The results of this parameter scan are shown in Figure~\ref{figstep2} and suggest a step size of $0.1$.

\section{The effect of the SBPS hyperparameters}
\label{MBandk}

In this section we explore, in the logistic regression example of Section 6.1, 
the effect of two hyperparameters that control the behavior of SBPS:
the mini-batch size $n$, and the width $k$ of the upper confidence band.
A third hyperparameter is the  rate of velocity refreshments,
shown in~\cite{bouchard2015bouncy} to be necessary in general to prove ergodicity.
But, as mentioned in Section~3, in the examples we considered the mini-batch noise 
was enough to sufficiently randomize possible non-mixing trajectories, so we could safely set this parameter to a very low value.

\subsection{Mini-batch size $n$}
\label{mbsize}
\begin{figure}[t!] 
  \centering
  \includegraphics[width=.9\textwidth]{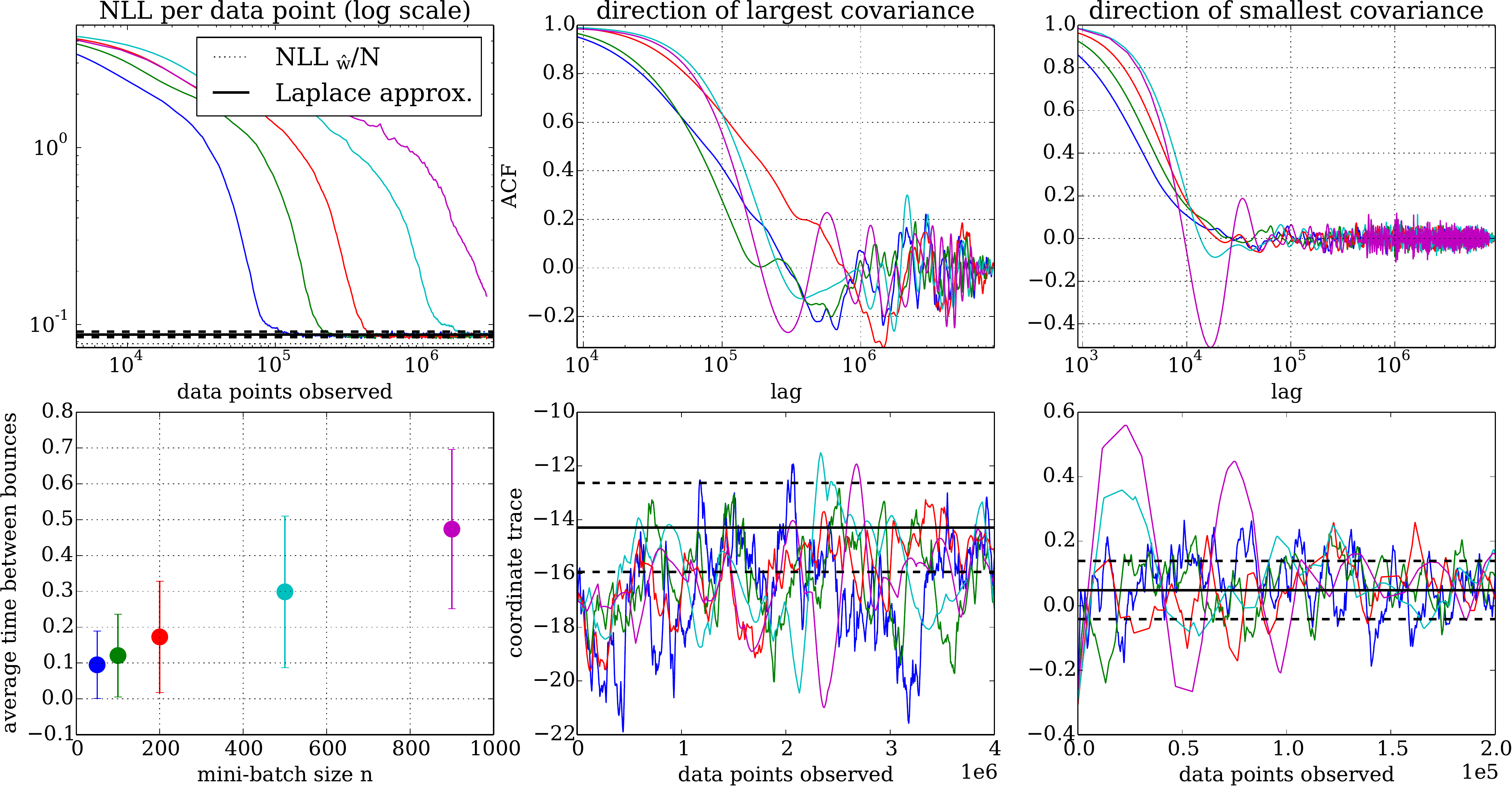}
     \caption{Effect of mini-batch sizes $n$ in the logistic regression example of Section 6.1. Mini-batch sizes were $50,100,200,500,900$.             
     \textit{Top Left:} Average per-data-point NLL over 5 runs. Note that smaller $n$ lead to faster convergence to a region of low NLL.
     \textit{Lower Left:} Estimated average time between particle bounces.      
     \textit{Center/Right:} ACF and trajectories from a single run, in the directions of smallest and biggest covariance. The x axis was chosen differently for the trajectory plots for clarity.}
\label{mbscan_fig}
\end{figure}

Figure~\ref{mbscan_fig} shows an exploration of different values of the mini-batch size $n$.
Low values for $n$ lead to high noise for $\tilde{G}$. This in turn yields higher values for the proposal intensity $\gamma(t)$, 
which leads to shorter linear trajectories between bounce proposals. 
This is consistent with the results of Figure~\ref{mbscan_fig} that show a linear relation between $n$
(i.e. computational cost per bounce proposal) and the average travel time between bounces. 
The autocorrelation functions (ACFs)  were computed from discrete samples obtained by 
running SBPS with different $n$'s such that the total data cost was the same for all cases, 
and then discretizing the continuous paths into equal numbers of uniformly spaced samples.
As shown, these cost adjusted ACFs are quite similar. 
On the other hand, the upper-left panel, shows that lower values of $n$ have faster convergence to equilibrium, suggesting 
that low $n$ should be preferred. But this should be contrasted with the fact that 
shorter linear trajectories increase the variance of expectations over rapidly changing functions, as discussed in Section~6.3. 


\begin{figure}[t!] 
  \centering
  \includegraphics[width=.9\textwidth]{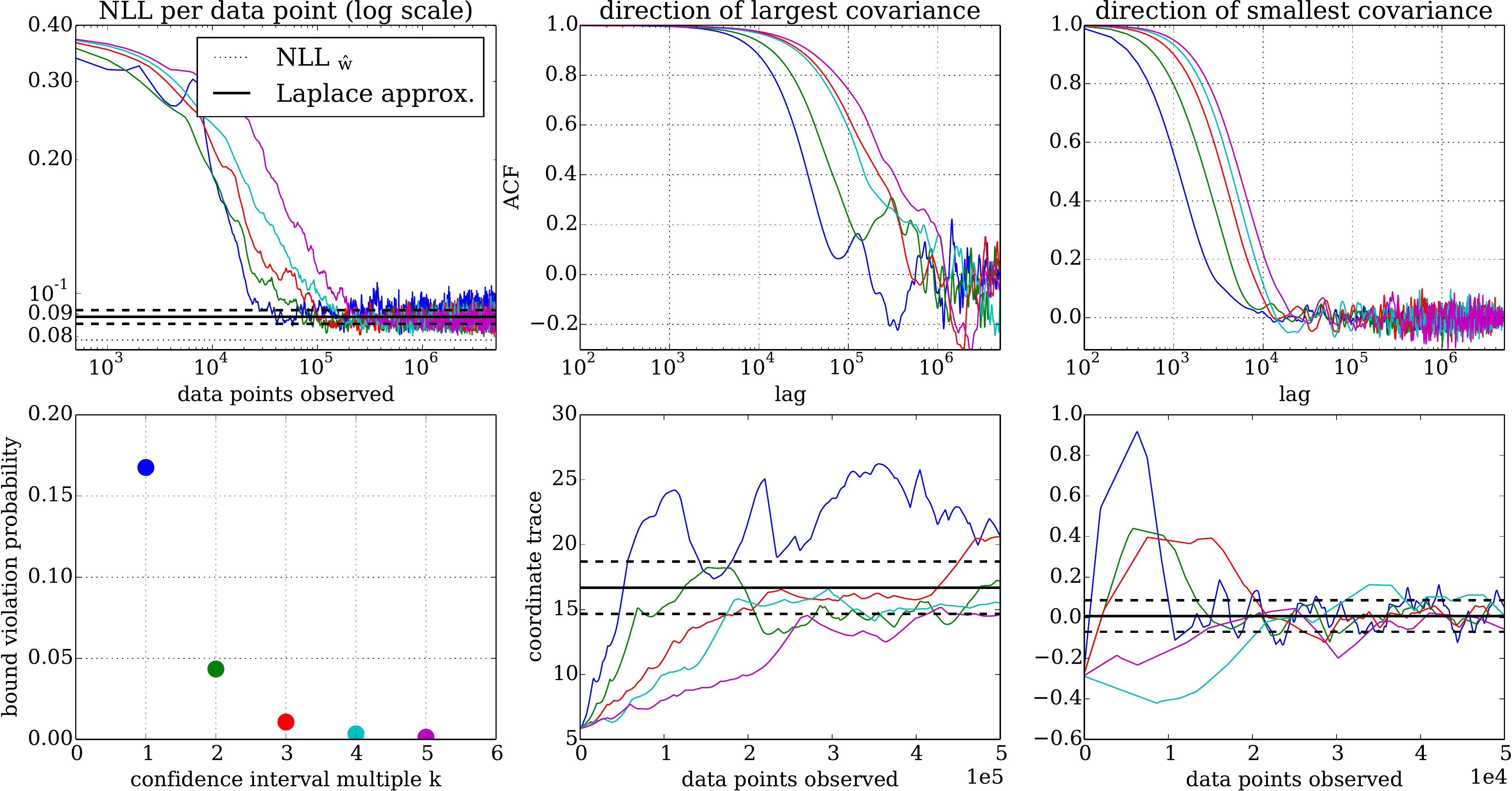}
     \caption{Effect of upper band size $k$ in the logistic regression example of Section 6.1, run with mini-batch size $n=100$.      
    \textit{Bottom Left:} Rate of upper bound violations as a function of $k$; the same colors are used in the other plots.           
     \textit{Top Left:} NLL per data point for samples of SBPS with different k values.   
     \textit{Center/Right:} ACF and trajectories in the directions of smallest and biggest covariance.      
     Note that smaller $k$ leads to faster convergence and mixing but increased bias, as visible in the coordinate trace in the direction of biggest covariance. The x axis was chosen differently for the trajectory plots for clarity.     
     }
\label{kscan_fig}
\end{figure}

\subsection{Upper-band width $k$}
Figure~\ref{kscan_fig} shows an exploration of different values of $k$, 
the height of the proposal intensity  above the estimator mean, in units of predictive standard deviation (see in Eq.(12) in main text).
It therefore controls the trade-off between a regime, at low $k$, of faster mixing and high bias from violations of the thinning upper bound ($[\tilde{G}(t)]_+/\lambda(t)>1$),  
 and another regime,  high $k$, of low bias and high variance from slower mixing.
As expected, the probability of bound violation decreases monotonically with $k$, as seen in the bottom left panel of Figure \ref{kscan_fig}.


\section{Upper Bounds for Logistic Regression}
\label{lip_bound}
In the case of logistic regression with data $(y_i, \x_i)$ the estimator of  $\nabla_\w U(\w)$ from a mini-batch of size $n$ is 
\eqan
\nabla_\w \tilde{U}(\w)=\frac{N}{n}\sum_{i=1}^n \x_i (\sigma(\w\cdot \x_i) -  y_i) \,.
\enan
A simple bound on $\tilde{G}(t)$ is therefore given by 
\begin{align}
\tilde{G}(t) &\leq \frac{N}{n} |\sum_{i=1}^n (\mathbf{v} \cdot \x_{i}) (\sigma(\w\cdot \x_i) -  y_i)| \,, \\
&\leq \frac{N}{n}\sum_{i=1}^n||\mathbf{v}||_2||(\sigma(\w\cdot \x_i) -  y_i)\x_{i}||_2 \,, \\
&\leq \frac{N}{n}\sum_{i=1}^n||x_{i}||_2 \,, \\
&\leq \sqrt{d} N \max_{i,j}| x_{ij} |  \,.
\end{align}
This is a particular case of a bound derived in~\cite{bierkens2017piecewise}.
Compared to the bound 
proposed in \cite{bouchard2015bouncy}, 
this bound is more conservative but cheaper to compute and does not require non-negative covariates.
It similarly  scales like $N$ and 
when the data used in the experiments was modified 
so that the covariates were non-negative the bounds differed by a factor lower than 2.

\bibliography{../thebib}
\bibliographystyle{icml2017}

\end{document}